\pdfoutput=1
\documentclass[11pt]{article}

\usepackage[preprint]{acl}

\usepackage{times}

\usepackage[T1]{fontenc}

\usepackage[utf8]{inputenc}

\usepackage{microtype}

\usepackage{inconsolata}

\usepackage{graphicx}
\usepackage{booktabs}
\usepackage{multirow}
\usepackage{amsmath}
\usepackage{amssymb}

\newcommand{\cset}{\mathcal{S}}

\title{Uncovering Audio-Text Alignment in Large Audio-Language Models through Cross-Modal Concept Neurons}

\author{
  Tzu-Quan Lin\textsuperscript{1,2}\thanks{Work done as a Visiting Research Student at CSTR, University of Edinburgh.},
  Chih-Kai Yang\textsuperscript{2}\thanks{These authors contributed equally as second authors.},
  Yi-Cheng Lin\textsuperscript{2}\textsuperscript{\dag},
  Hao Tang\textsuperscript{1},
  Hung-yi Lee\textsuperscript{2,3} \\
  \textsuperscript{1}Centre for Speech Technology Research, University of Edinburgh, United Kingdom \\
  \textsuperscript{2}Graduate Institute of Communication Engineering, National Taiwan University, Taiwan \\
  \textsuperscript{3}NTU Artificial Intelligence Center of Research Excellence (NTU AI-CoRE), Taiwan
}

\begin{document}
\maketitle
\begin{abstract}
While the language model backbone is pretrained on text alone,
large audio-language models (LALMs) can, to a great extent,
recognize acoustic attributes, such
as emotion, speaker gender, and spoken language.
However, it is unclear how the audio representation is integrated,
if at all, with the other text representation
before the model produces text.
We hypothesize that there is a shared subspace where
the audio and text representations are aligned.
To test this hypothesis, we identify cross-modal
neurons that are responsible for representing
both audio and text.
We find alignment between audio and text neurons to various degrees
across three LALMs and four acoustic attribute types.
Causal interventions, steering and deactivating the neurons,
confirm that the cross-modal neurons are not just necessary for prediction
but can also be manipulated.
\end{abstract}

\begin{figure}[t!]
  \centering
  \includegraphics[width=\columnwidth]{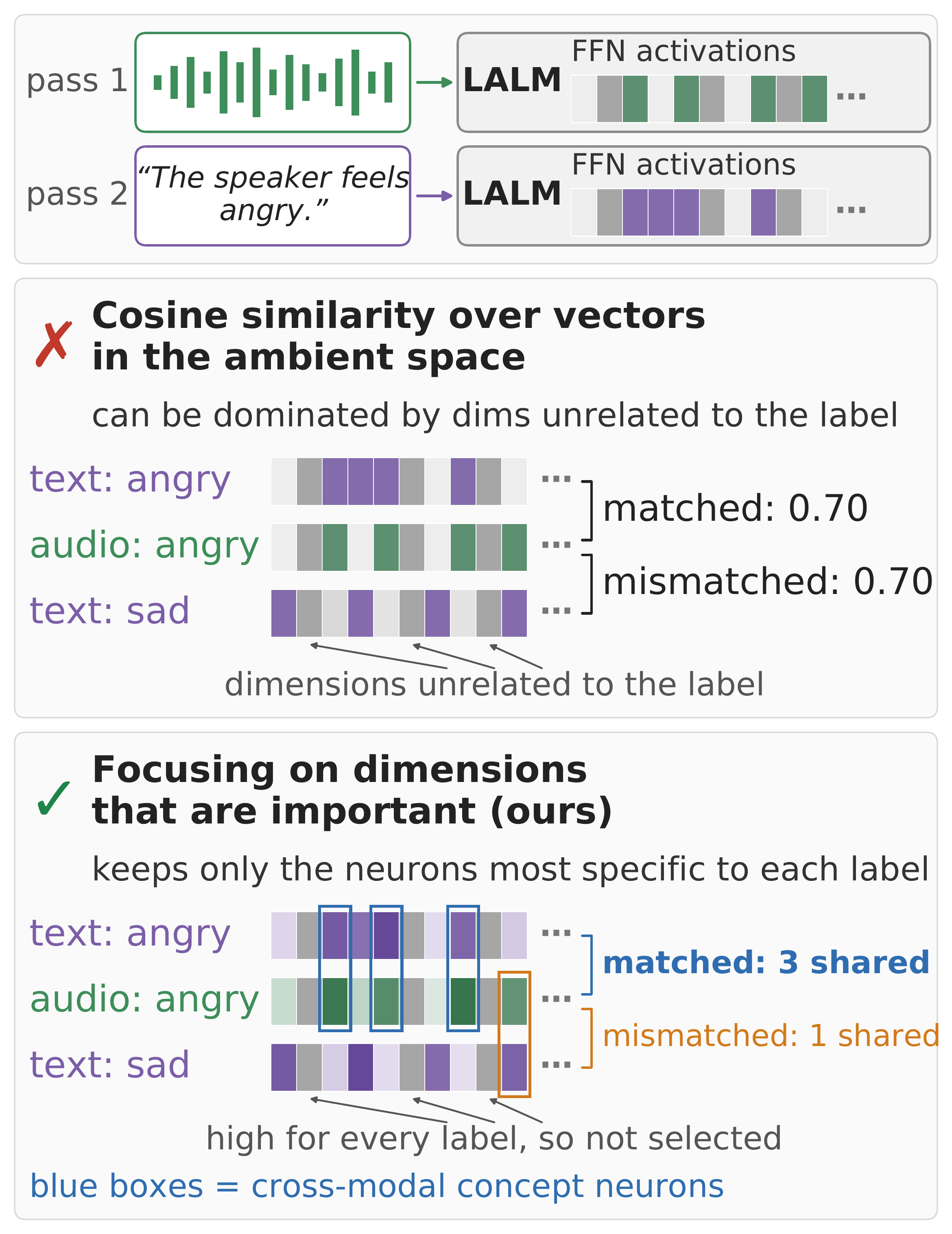}
  \caption{Why we compare neurons rather than entire vectors in the ambient space (schematic).
    Audio and text are fed to the LALM in separate forward passes.
    Cosine values follow Qwen2.5-Omni on emotion (\S\ref{sec:exp2});
    neuron counts are illustrative (\S\ref{sec:method}, \S\ref{sec:exp1}).}
  \label{fig:method_identification}
\end{figure}

\section{Introduction}
\label{sec:intro}

There are increasingly many large audio-language models (LALMs)
that process acoustic input directly through an LLM
backbone~\citep{chu2024qwen2audio,tang2024salmonn,gong2024ltu}, enabling
tasks such as speaker, emotion, and spoken-language identification from raw
audio~\citep{yang2024airbench,huang2024dynamicsuperb,huang2025dynamicsuperbphase2,sakshi2025mmau}.
Since the representations of an LALM's LLM backbone originate from
text pretraining, this raises a fundamental question: how does a backbone
whose representations were built for text come to recognize acoustic attributes?
It is less likely that the audio representation remains
separate from the text representation given how the LLM backbone is trained.
If the audio representation is integrated with the text representation,
there should exist a joint space where the audio and the text representations
are aligned.
In this work, we hypothesize that LALMs recognize acoustic attributes
through such an \emph{audio-text alignment}.

Interpretability work on LALMs has approached related questions
from several angles: tracking how attribute information evolves
across layers~\citep{yang2025audiolens,chen2026causaltracing},
locating attention heads that govern how much the model attends
to audio~\citep{glazer2026listening}, and identifying individual
neurons with attribute-specific function,
such as for emotion~\citep{zhao2026multilingualemotion,zhao2026emotionsensitive}.
One line of work shows that audio-language models project the linguistic
content of speech (what is being said) into a representation aligned
with the text embedding space~\citep{ogunremi2025transcribe}. These studies
either analyze the audio on its own or relate speech content,
rather than paralinguistic or speaker attributes, to text.
These results tell us little about whether and how audio representations
are integrated with text representations.

The work closest to answering our question is the Semantic Hub
hypothesis~\citep{wu2025semantichub}, which assumes that multimodal LLMs
route heterogeneous inputs, including acoustic signals such as animal
sounds, through a shared, text-anchored semantic space.
Their evidence, limited to a single LALM and to non-speech sounds,
rests on cosine similarity between vectors in the ambient space (i.e., the entire hidden vectors).

Most dimensions of vectors in the ambient space, however, carry information unrelated
to any particular label, such as neurons that are highly active for every
input~\citep{sun2024massive}.
A high cosine similarity does not show that the label itself
is shared, and a low one does not rule it out.
In fact, in our experiments with three LALMs (\S\ref{sec:exp2}), cosine similarity
barely tells matching from mismatching audio-text label pairs apart, and
label-independent dimensions account for $88\%$--$100\%$ of its variance.
A semantic hub, if it exists, should therefore be sought
in specific dimensions rather than across all dimensions of the representation.
We thus compare only a few dimensions in audio and in text,
those most relevant to each label, instead of the entire vectors.
This keeps the comparison from being dominated by dimensions
unrelated to the label and pinpoints the specific dimensions that carry
the alignment for causal interventions.

For these dimensions, we take
the individual neurons of the model's feed-forward
(FFN) layers, following the key-value memory view of the FFN
layer~\citep{geva2021keyvalue}. Prior work on self-supervised
speech models shows that such FFN dimensions already store
speech-related properties such as phonetic classes, speaker identity,
and pitch~\citep{lin2024propertyneurons,lin2025speakerneurons}.
Using the same approach, we identify neurons of certain
attributes for audio and for text.
If there exists a shared subspace, there should be a non-trivial
overlap between neurons for audio and for text, and we
refer to the overlapping neurons as \emph{cross-modal concept neurons}
(Figure~\ref{fig:method_identification}).

After identifying the cross-modal neurons, it is important to make sure
that they are necessary for predicting the audio attributes and
have a causal relationship with the concepts of interest.
We verify the importance of cross-modal neurons using two approaches
of causal intervention, steering~\citep{rimsky2024steering} and neuron deactivation (zeroing their activations)~\citep{dai2022knowledge}.

Across three LALMs (Qwen2.5-Omni, DeSTA2.5-Audio, and Audio Flamingo 3) and four attribute types (animal sound, emotion, gender, and spoken language), we find:
\begin{itemize}
    \item \textbf{The alignment exists but is hidden from cosine similarity.} Label-specific audio and text neurons coincide far more for matching than for mismatching labels, whereas cosine similarity barely tells matching from mismatching labels apart (\S\ref{sec:exp2}, \S\ref{sec:exp1}).
    \item \textbf{Interventions defined from text change recognition from audio.} Shifting cross-modal concept neurons along a text-derived direction raises the target answer significantly more than random neurons for 35 of 48 labels in Qwen2.5-Omni and Audio Flamingo 3 (7 of 24 in DeSTA2.5-Audio), and can make Qwen2.5-Omni report ``happy'' for every silent input ($20\%$ without steering; best random set $45\%$; \S\ref{sec:exp4}).
    \item \textbf{Recognition relies on cross-modal concept neurons.} Deactivating them impairs the target task before unrelated tasks in all 8 task combinations of Qwen2.5-Omni and Audio Flamingo 3 (2 of 4 in DeSTA2.5-Audio), with only $0.2$--$0.7\%$ of FFN neurons in Qwen2.5-Omni, and is as damaging as deactivating the most audio-specific neurons although they are less audio-specific (\S\ref{sec:exp3}). DeSTA2.5-Audio's weaker effects may partly reflect the ASR transcript it also receives.
\end{itemize}

\section{Cross-Modal Concept Neurons}
\label{sec:method}

To understand whether audio and text are mapped to a shared
subspace, we study audio recordings of a particular attribute paired with text descriptions of it.
Consider, for example, the emotion attribute \emph{angry}.
We run two forward passes through an LALM, one with
a speech utterance of an angry voice followed by
the question ``What is the emotional tone of the speaker
in the audio?'' and the other with only a text description of the label,
``The speaker feels angry'', i.e., the answer we would expect.
If there is a shared subspace for both audio
and text, then the two forward passes should produce
representations with a sufficient overlap.
For the rest of this section, we will describe
how we construct the audio and text representations
and how the overlap of the two is defined.

\paragraph{FFN key-value memories.} 
Following the key-value memory view of the Transformer
feed-forward (FFN) layer~\citep{geva2021keyvalue}, the first weight matrix
of an FFN layer defines a set of keys: each row is compared against a
token's residual-stream representation, and the resulting value, after
a nonlinearity, is what we call a neuron's activation, the strength
with which that token matches this particular key. The second FFN
weight matrix defines a corresponding set of values: each neuron's
activation scales how much of its associated value vector is written
back into the residual stream. Each neuron thus signals whether a fixed
direction is active, making neurons a fine-grained unit for testing
whether audio and text share such directions, as in neuron-level analyses
of text~\citep{geva2022transformer,dai2022knowledge} and speech
models~\citep{lin2024propertyneurons,lin2025speakerneurons}.

Formally, for an input $x$ and token
position $p$ at layer $\ell$ of an LALM decoder,
the output vector of the first FFN layer
(the SwiGLU output that feeds \texttt{down\_proj}) is denoted
as $h_\ell^p(x) \in \mathbb{R}^D$.
A neuron is simply a dimension in $h_\ell^p(x)$, indexed by $(\ell, i)$ with activation $h_\ell^p(x)[i]$.
We drop $\ell$ when it is clear from the context.

\paragraph{Pooling.}
Since the audio and text inputs span several tokens, 
we use the pooled hidden vectors from the first FFN layer
as the representation of the audio and the text.
For an input $x$ (either an audio clip paired with a question or a text description),
let $\mathcal{P}^{\text{audio}}(x)$ be the token positions
for the audio in the question (typically between the audio start
and end markers) and let $\mathcal{P}^{\text{text}}(x)$ be the 
token positions of the text description.
The representations are pooled at the instance level to
get the activations of a single instance,
and the overall activations are then
averaged across instances of the same attribute.
Formally, for a label $c$,
\begin{align}
\bar{h}^m(x) &= \frac{1}{|\mathcal{P}^m(x)|} \sum_{p \in \mathcal{P}^m(x)} h^{p}(x) \\
\mu^m(c) &= \frac{1}{|\mathcal{X}_c^m|} \sum_{x \in \mathcal{X}_c^m} \bar{h}^m(x)
\end{align}
where $m \in \{\text{text}, \text{audio}\}$,
$\mathcal{X}_c^\text{audio}$ is the set of audio clips paired with questions
that entail the label $c$ (for example, angry),
and $\mathcal{X}_c^\text{text}$ is the set of text descriptions
that entail the label $c$.

\paragraph{Distinctiveness.}
To find the neurons (i.e., individual dimensions),
we could rank the dimensions based on the average activations.
However, it is common that some neurons simply have a high activation baseline
across all labels~\citep{sun2024massive}.
To compensate for this, we subtract the baseline (based on other labels)
from the average activations and refer to this as the distinctiveness.
Formally, we compute the distinctiveness for label $c$ as
\begin{equation}
s^m(c) = \mu^m(c) - \frac{1}{|\mathcal{C}\setminus\{c\}|}
  \sum_{c' \in \mathcal{C}\setminus\{c\}} \mu^m(c'),
\end{equation}
where $m \in \{\text{text}, \text{audio}\}$ and $\mathcal{C}$ is the label set.
Each dimension of $s^m(c)$ is the distinctiveness of the corresponding neuron
for the modality $m$.

\paragraph{Cross-modal concept neurons.}
Finally, we select the top-$k$ neurons, denoted as
$\text{TopK}_k^m(c)$ for the modality $m$,
by ranking the distinctiveness of neurons together from all layers
$s_1^m(c), \dots, s_L^m(c)$ for an LALM decoder that has $L$ layers.
The intersection of the audio and text neurons
\begin{equation}
\cset(c) = \text{TopK}_k^{\text{audio}}(c) \cap \text{TopK}_k^{\text{text}}(c)
\end{equation}
is the set of cross-modal concept neurons of the label $c$.
The hyperparameter $k$ differs across experiments
and is described individually in later sections.
Comparing the two top-$k$ sets ensures that only dimensions
specific to label $c$ enter the comparison,
whereas cosine similarity over entire vectors includes every dimension.

\section{Experimental Setup}
\label{sec:setup}

We study three modern LALMs that combine an audio encoder with a text-pretrained LLM backbone: Qwen2.5-Omni-7B \citep{xu2025qwen25omni}, DeSTA2.5-Audio \citep{lu2025desta25audio}, and Audio Flamingo 3 \citep{goel2025audioflamingo3}. Qwen2.5-Omni and Audio Flamingo 3 update their backbone (Qwen2.5-7B) on audio-text data, whereas DeSTA2.5-Audio keeps its backbone (Llama-3.1-8B) frozen and aligns audio to it through a trained adapter. DeSTA2.5-Audio also feeds its backbone a Whisper transcript of the audio as text tokens whenever speech is detected; our audio-side activations (\S\ref{sec:method}) exclude these tokens. We return to these differences in \S\ref{sec:exp3}.

We draw our four attribute-recognition tasks from the SAKURA benchmark \citep{yang2025sakura}: emotion (5 labels: disgust, sad, angry, fear, happy), animal sound (9 labels: cat, cow, crow, dog, frog, hen, pig, rooster, sheep), gender (2 labels: female, male), and spoken language (8 labels: Chinese, English, French, German, Italian, Japanese, Korean, Spanish). We use all 500 audio instances of each task with SAKURA's single-hop multiple-choice questions. 
For every label, we write five paraphrased text descriptions (e.g., ``The speaker feels angry.'', ``This is an angry voice.''; full list in Appendix~\ref{sec:appendix_prompts}), so that the text side is not tied to one surface form. This audio-text pairing per label is what ``cross-modal'' refers to. 

\section{Experiments and Results}
\label{sec:results}

\subsection{Cosine Similarity Barely Reveals the Alignment}
\label{sec:exp2}

We first ask whether the alignment is visible in representational similarity, as used in prior work. For each label $c$, we concatenate its pooled per-layer FFN activations (\S\ref{sec:method}) across all layers, $\mu^m(c) = [\mu_1^m(c); \ldots; \mu_L^m(c)]$, separately for audio and text. Comparing every audio label's vector against every text label's vector with cosine similarity gives a label-by-label matrix whose diagonal holds matched pairs (e.g.\ audio ``dog'' vs.\ text ``dog'') and whose off-diagonal holds mismatched pairs (e.g.\ audio ``dog'' vs.\ text ``cat''). To quantify how clearly this matrix separates matched from mismatched pairs, we decompose its variance, as in a two-way analysis of variance~\citep{montgomery2017design}, into label-independent row and column offsets (a label similar to every label of the other modality) and a matched-label component, the extra similarity of the diagonal entries. The \emph{matched-label variance share} asks how much of the variation among the matrix's entries comes simply from whether the audio and text labels match, rather than from some labels being similar to everything (the offsets) or from noise. It is $100\%$ for a matrix that is a uniform bright diagonal on a flat background, and $0\%$ when matched pairs are no more similar than mismatched ones (formula in Appendix~\ref{sec:appendix_varshare}). Being a proportion, it can also be computed for the neuron-level overlap matrix of \S\ref{sec:exp1} and compared with the cosine share of the same model and task.

Table~\ref{tab:variance_share} reports this share for all 12 model-task combinations. For cosine similarity, the matched-label component is present but small: it accounts for less than $2\%$ of the variance for emotion, gender, and language in every model, and at most $9.7\%$ even for animal, while label-independent offsets account for $88\%$--$100\%$. Cosine similarity thus registers cross-modal alignment only as a small residual that is largely swamped by structure unrelated to label identity. The same decomposition applied to the neuron-level overlap matrices of \S\ref{sec:exp1} yields a share from matched labels of $9\%$--$92\%$, higher than the cosine share in every combination.

\begin{table}[t!]
  \centering
  \footnotesize
  \begin{tabular}{llrr}
    \toprule
    \textbf{Model} & \textbf{Task} & \textbf{Cosine} & \textbf{Neuron} \\
    \midrule
    Qwen2.5-Omni     & animal   & 8.74\% & 70.08\% \\
                     & emotion  & 0.88\% & 25.04\% \\
                     & gender   & 0.16\% & 76.04\% \\
                     & language & 0.23\% & 55.08\% \\
    \midrule
    DeSTA2.5-Audio   & animal   & 0.07\% & 9.22\% \\
                     & emotion  & 1.58\% & 52.03\% \\
                     & gender   & 1.85\% & 91.69\% \\
                     & language & 0.06\% & 9.73\% \\
    \midrule
    Audio Flamingo 3 & animal   & 9.69\% & 67.05\% \\
                     & emotion  & 1.05\% & 45.46\% \\
                     & gender   & $<$0.01\% & 68.57\% \\
                     & language & 0.38\% & 59.30\% \\
    \bottomrule
  \end{tabular}
  \caption{
Matched-label variance share (\S\ref{sec:exp2}): how much of the variation in each audio-by-text label matrix comes from whether the labels match, rather than from labels similar to everything or from noise. Cosine: cosine similarity of pooled activations; Neuron: Jaccard overlap of cross-modal concept neuron sets (\S\ref{sec:exp1}).} 
  \label{tab:variance_share}
\end{table}

\subsection{Neurons Reveal a Label-Specific Audio-Text Alignment}
\label{sec:exp1}

We next ask whether the alignment is visible at the level of individual neurons, testing two properties of $\cset(c)$ (\S\ref{sec:method}). Here, $k$ is tuned per model and task on a held-out validation split (20\%) over the same grid $\{25, 50, 100, 200, 500, 1000, 2000, 4000\}$ for every combination, maximizing mean per-label Jaccard overlap~\citep{jaccard1912distribution}; the selected $k$ is 100 or 200 in all 12 combinations, and with it we reselect the neuron sets on the held-out test split, on which all overlap results are reported.

\emph{Existence}: as a basic precondition, we first check that $\text{TopK}_k^{\text{audio}}(c)$ and $\text{TopK}_k^{\text{text}}(c)$ overlap more than two unrelated neuron sets of the same size would, measured by their Jaccard overlap $J(A,B) = |A\cap B| / |A\cup B|$, tested per label with an exact hypergeometric test~\citep{rivals2007enrichment}; the largest per-label $p$-value over all 12 combinations is below $10^{-29}$ (Appendix~\ref{sec:appendix_hypergeom}; Table~\ref{tab:appendix_exist_p}). 

\emph{Distinctiveness}: whether this overlap is particular to the correct label pairing rather than a generic property of any audio-text pair, tested by reusing the same top-$k$ sets from the existence test above in a label-by-label matrix construction analogous to \S\ref{sec:exp2}'s, with Jaccard overlap in place of cosine similarity, and tested with an exact label-permutation test~\citep{mantel1967detection} on the diagonal-vs-off-diagonal gap (full procedure in Appendix~\ref{sec:appendix_perm}); Table~\ref{tab:exp1_existence} marks the resulting significance levels. Matched audio-text label pairs share substantially more neurons than mismatched pairs, significant in all nine testable model-task combinations (animal, emotion, language). Gender, with only two labels, cannot be tested by label permutation (its smallest attainable $p$-value is $0.5$). Descriptively, its matched audio-text pairs nonetheless share more neurons than its mismatched pairs in every model, with diagonal-to-off-diagonal ratios of mean Jaccard overlap ($1.35$--$2.77\times$; Table~\ref{tab:appendix_ratio} in Appendix~\ref{sec:appendix_perm}) comparable to those of the testable tasks. 
The alignment that cosine similarity barely registers is thus clearly visible at the neuron level, label by label, in all three models and all four attributes (matrices: Appendix~\ref{sec:appendix_exp2matrix}). 

\begin{table}[t!]
  \centering
  \footnotesize
  \setlength{\tabcolsep}{4pt}
  \begin{tabular}{lcccc}
    \toprule
    \textbf{Model} & \textbf{Anim.} & \textbf{Emo.} & \textbf{Gen.} & \textbf{Lang.} \\
    \midrule
    Qwen2.5-Omni     & 0.245$^{**}$ & 0.174$^{*}$  & 0.144$^{\dagger}$ & 0.177$^{**}$ \\
    DeSTA2.5-Audio   & 0.094$^{**}$  & 0.121$^{**}$ & 0.127$^{\dagger}$ & 0.125$^{**}$ \\
    Audio Flamingo 3 & 0.264$^{**}$ & 0.202$^{**}$ & 0.098$^{\dagger}$ & 0.216$^{**}$ \\
    \bottomrule
  \end{tabular}
  \caption{Cross-modal concept neurons (\S\ref{sec:exp1}): mean per-label Jaccard overlap between the audio and text top-$k$ sets on the held-out test split. 
Stars: distinctiveness, exact label-permutation test of the matched-vs-mismatched gap (Appendix~\ref{sec:appendix_perm}); $^{*}p<0.05$, $^{**}p<0.01$; $^{\dagger}$not testable (gender has only two labels, so its smallest attainable $p$ is $0.5$). Exact values: Table~\ref{tab:appendix_exist_p}.}
  \label{tab:exp1_existence}
\end{table}

\begin{figure*}[t]
  \centering
  \includegraphics[width=\textwidth]{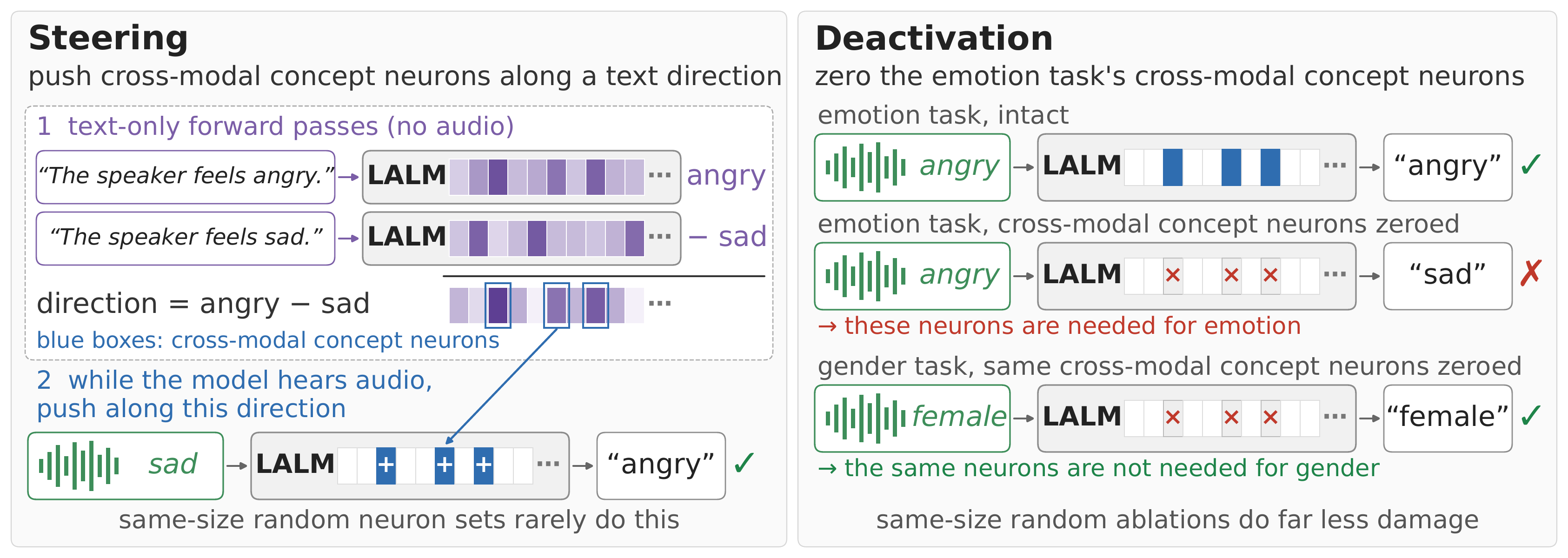}
  \caption{Overview of the causal tests (schematic; activation values are illustrative). Steering (\S\ref{sec:exp4}): the direction is computed once from text-only forward passes as the difference between the target and carrier labels' activations, and is added to the target label's cross-modal concept neurons while the model processes a carrier input (Eq.~\ref{eq:patch}); the carrier can also be silence or noise. Deactivation (\S\ref{sec:exp3}): deactivating a task's cross-modal concept neurons (Eq.~\ref{eq:ablation}) impairs that task before the other tasks, whereas deactivating a random set of the same size does far less damage.}
  \label{fig:method_causal}
\end{figure*}

\subsection{Text-Derived Interventions Steer Audio Recognition}
\label{sec:exp4}

If audio and text are aligned in these neurons, an intervention defined from text alone should change what the model recognizes from audio. We test this with \emph{contrastive neuron steering} (Figure~\ref{fig:method_causal}) \citep{turner2023activation,rimsky2024steering}, which shifts the cross-modal concept neurons along a text-derived direction. With reference vector $v^{\text{text}}(c) = \mu^{\text{text}}(c)$ (\S\ref{sec:method}'s mean-of-means, concatenated across layers) and direction $\Delta^{\text{text}}(c_t, c_r) = v^{\text{text}}(c_t) - v^{\text{text}}(c_r)$ between a target label $c_t$ and a same-task \emph{carrier} label $c_r$, we shift
\begin{equation}
\tilde h_\ell[i] = h_\ell[i] + \lambda \cdot \Delta_\ell^{\text{text}}[i](c_t, c_r), \quad (\ell,i) \in \cset(c_t)
\label{eq:patch}
\end{equation}
at every token position of \texttt{down\_proj}'s input, while the model processes an audio instance of carrier $c_r$. 
Shifting toward a value the network has genuinely produced elsewhere, rather than by an arbitrary multiple of its current activation, keeps this closer to the training distribution even at large coefficients~\citep{zhang2024best}.
 We sweep $\lambda \in \{0.5, 1, 1.5, 2, 3, 5\}$ and measure the \emph{target rate}: the fraction of samples whose predicted answer is the target label.  Since $\Delta^{\text{text}}$ comes from text alone, a rise in the target rate means that a change defined in text transfers to recognition from audio. This holds even when steering only audio-token positions (Appendix~\ref{sec:appendix_audiopos}).

Every other label in the task with test samples whose answer options include $c_t$ (up to 20 per label) serves as a carrier $c_r$, plus two non-speech synthetic carriers (silence, noise; Appendix~\ref{sec:appendix_synthcarrier}), so no single carrier drives an observed effect. We compare the label's own cross-modal concept neurons $\cset(c_t)$ against five independent \emph{random} neuron sets of the same size $N$ (the null), each drawn uniformly from all FFN neurons outside $\cset(c_t)$ and shared across all carriers of the label; $k$ (and hence $N$) is chosen as described in Appendix~\ref{sec:appendix_ksuff}. 
We summarize each neuron set by its target rate averaged over carriers and $\lambda$, and test whether the cross-modal concept neuron set lies above the five random sets per label with a one-sided prediction-interval $t$-test~\citep{hahn1991statistical}, with Benjamini--Hochberg correction across all 72 labels~\citep{benjamini1995controlling}, and per model with a rank test~\citep{friedman1937use} that asks how likely so low a rank sum, over the model's labels, would be if the cross-modal concept neuron set were just another random set (Appendix~\ref{sec:appendix_suff}). 

Table~\ref{tab:exp4_sufficiency} summarizes the results. Cross-modal concept neurons produce a significantly higher target rate than all five random sets in 42 of the 72 model-task-label combinations. The effect is clearly model-dependent: 18 of 24 labels reach significance in Qwen2.5-Omni and 17 of 24 in Audio Flamingo 3, but only 7 of 24 in DeSTA2.5-Audio. 
The rank test agrees: the cross-modal concept neuron set's average rank among the six sets is $1.21$ (Qwen2.5-Omni) and $1.17$ (Audio Flamingo 3) against a null of $3.5$ ($p<10^{-13}$), but $2.50$ in DeSTA2.5-Audio ($p=0.002$): better than chance, but far less consistent. Steering DeSTA2.5-Audio is weakest for spoken language (1 of 8 labels), possibly because its transcript input, written in the spoken language (\S\ref{sec:setup}; Appendix~\ref{sec:appendix_desta_asr}), keeps revealing the carrier language during steering. 
 Because cross-modal concept neurons concentrate in later layers, we also compare against random sets with the same number of neurons per layer: on Qwen2.5-Omni's emotion task, the cross-modal concept neuron set still beats all five for every label (Appendix~\ref{sec:appendix_layermatched}). 

\begin{figure*}[t]
  \centering
  \includegraphics[width=0.8\textwidth]{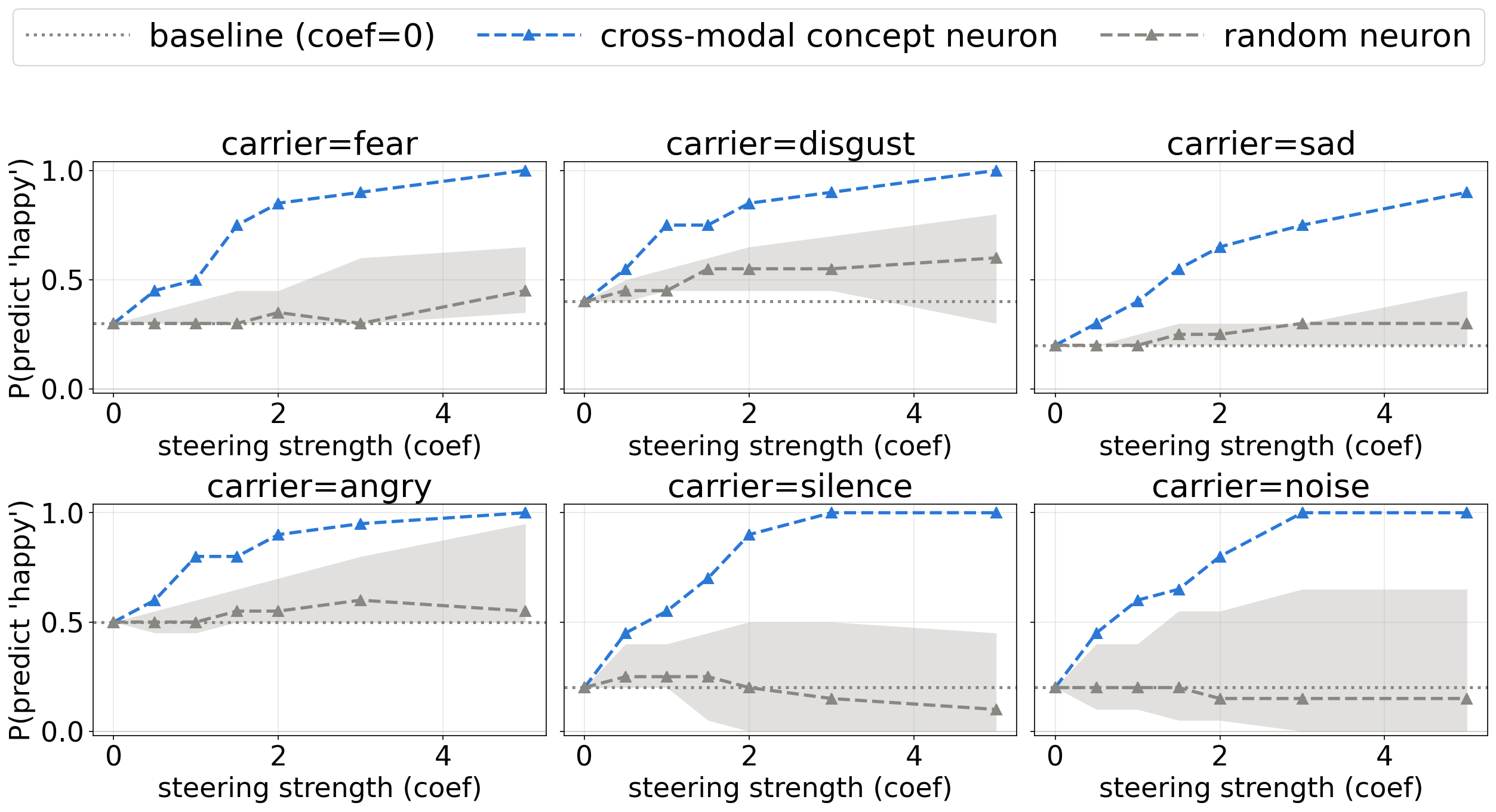}
  \caption{Contrastive neuron steering toward the emotion label ``happy'' in Qwen2.5-Omni, one panel per carrier label. $x$-axis: steering coefficient $\lambda$; $y$-axis: fraction of samples predicted as ``happy''. Blue: cross-modal concept neurons; grey: median of five matched-size random neuron sets, with their min--max range shaded (the null).}
  \label{fig:exp4_qwen_happy}
\end{figure*}

 Figure~\ref{fig:exp4_qwen_happy} shows one of the strongest cases, Qwen2.5-Omni's ``happy'': the cross-modal concept neurons push the target rate above all five random sets for every carrier, even silence and noise, which carry no acoustic evidence for any label (other models: Appendix~\ref{sec:appendix_exp4}).

\begin{table}[t!]
  \centering
  \footnotesize
  \setlength{\tabcolsep}{3.5pt}
  \begin{tabular}{lccccc}
    \toprule
    \textbf{Model} & \textbf{Anim.} & \textbf{Emo.} & \textbf{Gen.} & \textbf{Lang.} & \textbf{Total} \\
    \midrule
    Qwen2.5-Omni     & 4/9  & 5/5   & 2/2 & 7/8   & \textbf{18/24} \\
    DeSTA2.5-Audio   & 2/9  & 4/5   & 0/2 & 1/8   & \textbf{7/24} \\
    Audio Flamingo 3 & 6/9  & 4/5   & 1/2 & 6/8   & \textbf{17/24} \\
    \midrule
    Total            & 12/27 & 13/15 & 3/6 & 14/24 & \textbf{42/72} \\
    \bottomrule
  \end{tabular}
  \caption{Number of labels for which steering cross-modal concept neurons yields a significantly higher target rate than five matched-size random neuron sets (one-sided prediction-interval $t$-test, $q<0.05$ after Benjamini--Hochberg correction across all 72 labels; Appendix~\ref{sec:appendix_suff}).}
  \label{tab:exp4_sufficiency}
\end{table}

\subsection{Recognition Relies on Cross-Modal Concept Neurons}
\label{sec:exp3}

Finally, we ask whether recognition relies on the cross-modal concept neurons that carry the alignment: if their overlap were incidental, removing them should not selectively impair recognition. We compute each label's $\cset(c)$ using only the validation split (\S\ref{sec:method}), so the deactivated neurons are never selected on the same samples we then evaluate, and deactivate them by setting their activations to zero (zero-ablation; schematic in Figure~\ref{fig:method_causal}) with a forward pre-hook
\begin{equation}
\tilde h_\ell[i] = \begin{cases} 0 & (\ell,i) \in \cset \\ h_\ell[i] & \text{otherwise} \end{cases}
\label{eq:ablation}
\end{equation}
applied to \texttt{down\_proj}'s input at every token position, 
then measure accuracy on the task's SAKURA multiple-choice questions, where the model answers with a single option letter (parsing details in Appendix~\ref{sec:appendix_parsing}). 

We deactivate at the \emph{task level}: $\cset = \bigcup_{c \in \mathcal{C}} \cset(c)$, the union across a task's labels, and we compare accuracy on that task (target) to the other three (controls), to isolate an effect specific to the deactivated task from a generic disruption. We include five matched-size random-neuron control trials, plus a matched-size audio-only reference control: intuitively, deactivating the neurons most responsive to audio should be highly damaging, making it a natural reference point for checking whether cross-modal alignment specifically, rather than generic audio-responsiveness, drives any effect.

Since a small, fixed neuron-set size can understate a real effect, we repeat every deactivation for each $k$ in the grid of \S\ref{sec:exp1} extended by doubling up to $256{,}000$ (14 values), which deactivates $N=|\cset|$ neurons, from $2$ to about $500{,}000$ (Appendix~\ref{sec:appendix_nablated}), and summarize each resulting curve over $N$ by its \emph{decline onset}: the smallest $N$ at which accuracy drops by more than 10\% relative to baseline and stays down for the rest of the sweep, so that a brief dip that later recovers does not count. A combination is \emph{selective} if the target's onset precedes the controls'.

\begin{figure*}[t]
  \centering
  \includegraphics[width=\textwidth]{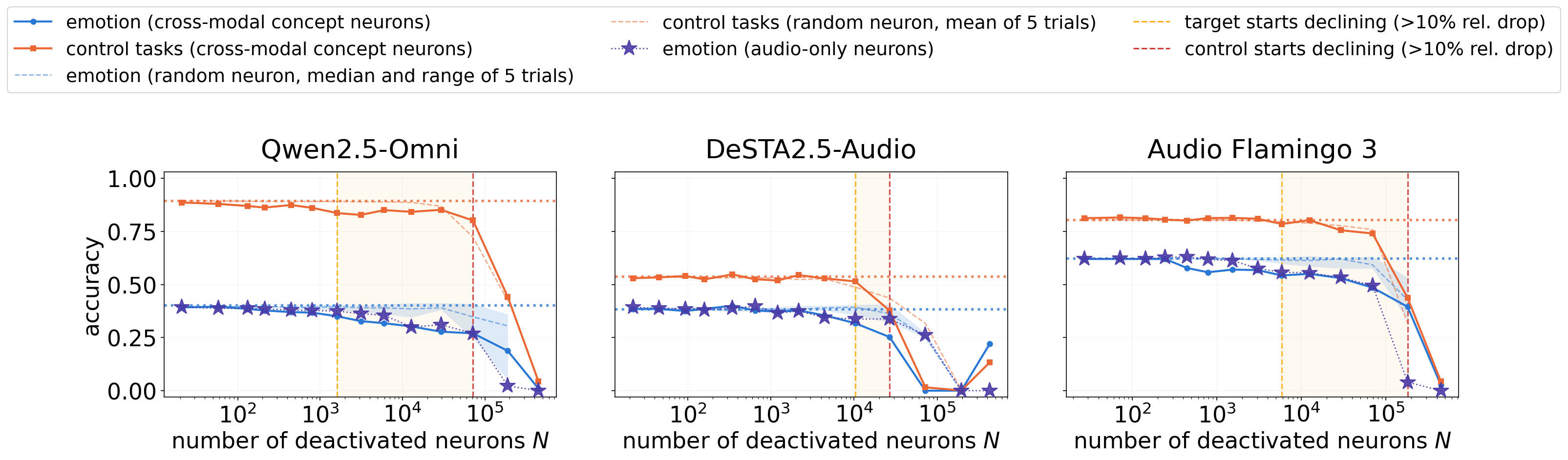}
  \caption{Task-level deactivation on the emotion task. Solid lines: target task accuracy (blue) and mean control-task accuracy (orange) versus the number of deactivated neurons $N$ (log scale). Dashed lines: matched-size random-neuron control (target task: median of the five random trials, with their min--max range shaded; control tasks: mean). Purple stars: audio-only neurons, a reference control (\S\ref{sec:exp3}). Vertical dashed lines mark the target's and controls' decline onsets, shaded when the target's comes first.}
  \label{fig:exp3_emotion}
\end{figure*}

Table~\ref{tab:exp3_necessity} reports this comparison for all 12 model-task combinations. Ten of the twelve show a selective necessity window: deactivating a task's own cross-modal concept neurons degrades that task's own accuracy before an equally sized deactivation degrades any unrelated task.  Deactivating them is also about as damaging as deactivating the audio-only reference set, although they are far less audio-specific: at each task's decline onset, their median rank in the audio specificity ranking is $1.5$--$5.0\times$ that of the audio-only neurons, with $37$--$74\%$ ranking below every audio-only neuron. Recognition thus relies on the neurons that carry the alignment, not merely on the most audio-specific ones. 
The result holds under a bootstrap over test samples (9 of the 10 windows selective in $\ge 90\%$ of resamples, Audio Flamingo 3's gender window in only $39\%$; Appendix~\ref{sec:appendix_bootstrap}) and deactivation restricted to audio-token positions (Qwen2.5-Omni emotion; Appendix~\ref{sec:appendix_audiopos}). Text descriptions using synonyms instead of the label word (e.g., ``The speaker feels furious.'' for \emph{angry}; Appendix~\ref{sec:appendix_synonym}) select largely the same neurons and give a nearly identical deactivation effect, so these neurons track the attribute rather than the word itself.  
 Figure~\ref{fig:exp3_emotion} shows the emotion curves: in all three models, target accuracy drops before control accuracy, while matched-size random-neuron deactivations decline only at much larger $N$, so the effect depends on which neurons are deactivated, not merely on how many (other tasks: Appendix~\ref{sec:appendix_exp3}). 

 The two non-selective combinations in Table~\ref{tab:exp3_necessity} are both DeSTA2.5-Audio, on the animal and language tasks, where the target task's accuracy declines at exactly the same $N$ as the controls', showing no evidence of selective necessity for these two tasks. As in \S\ref{sec:exp4}, its transcript input can reveal the spoken language without audio-driven neurons, but this does not explain the animal task, whose clips almost never receive a transcript; there, its frozen backbone or different backbone family may matter, which three models cannot separate.

\begin{table}[t!]
  \centering
  \footnotesize
  \setlength{\tabcolsep}{3pt}
  \begin{tabular}{llccc}
    \toprule
    \textbf{Task} & \textbf{Model} & \textbf{Target $N$} & \textbf{Control $N$} & \textbf{Sel.} \\
    \midrule
    \multirow{3}{*}{animal} & Qwen2.5-Omni      & 1{,}745   & 36{,}014  & yes \\
                           & DeSTA2.5-Audio    & 5{,}973   & 5{,}973   & no \\
                           & Audio Flamingo 3  & 984       & 123{,}275  & yes \\
    \midrule
    \multirow{3}{*}{emotion} & Qwen2.5-Omni      & 1{,}618   & 71{,}812  & yes \\
                           & DeSTA2.5-Audio    & 10{,}336  & 26{,}763  & yes \\
                           & Audio Flamingo 3  & 5{,}862   & 180{,}203 & yes \\
    \midrule
    \multirow{3}{*}{gender} & Qwen2.5-Omni      & 1{,}255   & 264{,}115 & yes \\
                           & DeSTA2.5-Audio    & 317   & 93{,}948 & yes \\
                           & Audio Flamingo 3  & 74{,}073 & 251{,}664 & yes \\
    \midrule
    \multirow{3}{*}{language} & Qwen2.5-Omni      & 3{,}817   & 107{,}461  & yes \\
                           & DeSTA2.5-Audio    & 5{,}430   & 5{,}430   & no \\
                           & Audio Flamingo 3  & 19{,}538  & 102{,}782  & yes \\
    \bottomrule
  \end{tabular}
  \caption{Task-level deactivation (\S\ref{sec:exp3}). The number of deactivated neurons $N$ at which target and control accuracy each begin to decline (more than 10\% relative to baseline, sustained for the rest of the sweep), and whether the target's onset precedes the controls' (selective necessity window).}
  \label{tab:exp3_necessity}
\end{table}

\section{Related Work}

\paragraph{Interpretability of vision-language models.} A parallel line of work studies how multimodal LLMs that pair a vision encoder with a text-pretrained backbone process visual information internally. \citet{goh2021multimodal} first showed that individual neurons in CLIP fire for the same concept whether it appears as an image or as text, coining the term ``multimodal neuron.'' \citet{schwettmann2023multimodal} found similar neurons inside a frozen, text-only LLM backbone augmented with a vision encoder: FFN neurons that translate projected visual features into the corresponding text concept in the residual stream, with an ablation showing a causal effect on captioning. None of this work looks at audio, or steers these neurons via activation intervention, the question central to \S\ref{sec:exp4}.

\paragraph{Interpretability of audio-language models.} Beyond the layer-, attention-head-, and neuron-level analyses discussed in \S\ref{sec:intro}, recent work has tested whether a spoken answer becomes ``readable'' from a middle layer before the model commits to it \citep{fan2026readmind}. Other work traced which internal pathway wins when audio and text evidence conflict, along the way reporting, via cosine similarity, that the two converge in later layers \citep{cho2026whowins}. None of these studies identifies individually shared audio-and-text units and tests both whether steering them changes the model's behavior and whether it relies on them.

\paragraph{Semantic Hub hypothesis.} 

Beyond audio, \citet{wu2025semantichub} also study multilingual models, where inputs in different languages share representations anchored in the dominant training language. Their evidence rests on cosine similarity over vectors in the ambient space, not on individual neurons (\S\ref{sec:intro}), and their causal test adds the difference between two word unembeddings at one layer, a steering test only; we additionally test deactivation, in three LALMs and on both speech and non-speech attributes. 

\paragraph{Shared units across surface forms.} Representing one concept with a single unit shared across different surface forms has been tested causally elsewhere: \citet{brinkmann2025shared} found that ablating a shared unit knocks out a grammatical concept across many typologically distinct languages while a language-specific unit barely matters. \citet{nakai2026generation} focus on neurons that control the \emph{output} language of speech--text models: they identify language-selective neurons separately from speech and text inputs and steer generation through the output-language neurons, whereas the neurons that represent the \emph{input} language across modalities are compared only by overlap, without testing whether the model relies on them. They also study language identity only, whereas we also cover paralinguistic and non-speech attributes.

\section{Conclusion}

LALMs recognize acoustic attributes through an alignment between audio and text that cosine similarity over entire representations barely reveals but individual FFN neurons expose: the neurons most specific to an attribute in audio and in text coincide, label by label. This alignment is truly cross-modal, since a direction computed from text steers what the model recognizes from audio (\S\ref{sec:exp4}), and recognition relies on it, since removing cross-modal concept neurons selectively impairs the attribute (\S\ref{sec:exp3}). How strongly a model relies on the alignment varies across models.

\section*{Limitations}

Our attributes and materials come entirely from the SAKURA benchmark: four fixed attribute sets and text descriptions we authored ourselves (\S\ref{sec:setup}); each label's text side consists of five templated paraphrases. A much larger prompt set per label is hard to construct without introducing label-specific wording that would itself confound the comparison across labels, so we instead use templates shared by every label of a task that differ only in the label word. 

Three LALMs is not an exhaustive sample of current architectures. In particular, DeSTA2.5-Audio differs from the other two models at once in its frozen backbone, its backbone family, and its transcript input, so explaining its weaker effects beyond the transcript route for spoken language would require more models.

The gender task follows SAKURA's binary female/male labels, which do not capture the full range of gender identities or voice characteristics; our gender results should be read as results on this binary recognition task.

Gender has only two labels, so the label-permutation test of \S\ref{sec:exp1} cannot assess its distinctiveness (the smallest attainable $p$-value is $0.5$), and we describe it only descriptively; the existence test and the random-set tests of \S\ref{sec:exp4} do not depend on the number of labels and do cover gender.

For the steering experiment (\S\ref{sec:exp4}), each label's neuron-set size is taken from its necessity window (\S\ref{sec:exp3}), which is located on the test split. We did not locate these windows on the validation split, which holds as few as 8 samples per label, too few to identify a sustained 10\% accuracy decline reliably. The selection uses a different intervention and outcome (deactivation, task accuracy) from those of the steering test (steering, target rate), so it does not directly optimize the steering result, and every random control set is matched to the same neuron-set size.

\section*{Ethical Considerations}

This work is purely diagnostic: we analyze the internal representations of three existing, publicly released pretrained LALMs using an existing public benchmark (SAKURA), and we neither collect new data, train new models, nor release any new model weights or fine-tuned checkpoint. The four attributes we study, animal-sound, emotion, gender, and spoken-language identification, are recognition capabilities already present in each model's public release; our steering and deactivation interventions (\S\ref{sec:exp4}, \S\ref{sec:exp3}) are used strictly as scientific probes to test whether specific neurons can steer, or are necessary for, this existing behavior, not as a proposed technique for modifying or improving deployed systems. We therefore do not foresee this work introducing risks beyond those already inherent in the publicly released models and benchmark it studies.

\section*{Acknowledgments}

Tzu-Quan Lin is supported by a joint Ph.D. scholarship from NTU GICE and Delta Electronics, Inc. This work was supported by the Ministry of Education (MOE) of Taiwan under the project Taiwan Centers of Excellence in Artificial Intelligence, through the NTU Artificial Intelligence Center of Research Excellence (NTU AI-CoRE).

We used an AI writing assistant (Claude) to help edit the grammar and prose of this manuscript. The AI assistant did not design any experiment, generate or select any result, or make any research decision; it did not contribute to the paper's scientific content. The authors are fully responsible for, and vouch for the accuracy of, all scientific content, experiments, and conclusions reported in this paper.

\bibliography{custom}

\appendix
\raggedbottom

\section{Significance Testing Details}
\label{sec:appendix_significance}

This appendix gives the full procedure behind each significance test used in \S\ref{sec:results}, which the main text describes only briefly.

\subsection{Matched-Label Variance Share}\label{sec:appendix_varshare}
For an $|\mathcal{C}|\times|\mathcal{C}|$ label-by-label matrix $M$ with row means $\bar r_i$, column means $\bar c_j$, and grand mean $\bar g$, we fit $M_{ij} = \alpha_i + \beta_j + \delta\,[i=j] + \varepsilon_{ij}$ by least squares, as in a two-way analysis of variance~\citep{montgomery2017design}. The row and column terms alone explain $SS_{\text{row+col}} = |\mathcal{C}|\sum_i(\bar r_i-\bar g)^2 + |\mathcal{C}|\sum_j(\bar c_j-\bar g)^2$ of the total $SS_{\text{total}} = \sum_{i,j}(M_{ij}-\bar g)^2$. Adding the diagonal term gives $\hat\delta = d$, the diagonal-minus-off-diagonal gap used by the permutation test below, and explains an additional $(|\mathcal{C}|-1)\,d^2$. The variance share from matched labels is $(|\mathcal{C}|-1)\,d^2 / SS_{\text{total}}$, and the row-and-column share is $SS_{\text{row+col}}/SS_{\text{total}}$. It is a descriptive effect size rather than a test; since it is a proportion, it can be compared directly between the cosine and neuron-overlap matrices of the same model and task. For gender ($|\mathcal{C}|=2$) the diagonal term is the only component beyond the row and column offsets.

\subsection{Hypergeometric Existence Test}\label{sec:appendix_hypergeom}
For a single label $c$, we ask whether $|\text{TopK}_k^{\text{audio}}(c) \cap \text{TopK}_k^{\text{text}}(c)|$ is larger than expected if the two size-$k$ sets were independent draws from the full population of $LD$ neurons. This probability is given exactly by the hypergeometric distribution~\citep{rivals2007enrichment}, with no permutation or approximation needed. Rather than combining per-label $p$-values across a task's labels, for instance with Fisher's method~\citep{fisher1925statistical}, which assumes independent $p$-values although all labels of a task are scored from the same activations, we report the largest per-label $p$-value of each task, which bounds every label's $p$-value without any independence assumption. The random baseline reported alongside each existence result (Table~\ref{tab:appendix_exist_p}'s ``Random baseline'' column) is the expected Jaccard overlap of two independent size-$k$ draws, estimated from 100,000 Monte Carlo samples rather than a closed-form expression.

\subsection{Diagonal-vs-Off-Diagonal Permutation Test}\label{sec:appendix_perm}
Given the label-by-label neuron-Jaccard overlap matrix of \S\ref{sec:exp1}, we take the observed gap between the mean diagonal (matched-label) entry and the mean off-diagonal (mismatched-label) entry. The null distribution comes from permuting the text-side label ordering~\citep{mantel1967detection}, enumerating all $|\mathcal{C}|!$ permutations exactly for a task with $|\mathcal{C}|$ labels ($|\mathcal{C}| \le 9$), and recomputing this gap for each, so that under the null, any audio label is equally likely to be paired with any text label. The one-sided $p$-value is the fraction of permutations (including the identity) whose gap is at least as large as the observed gap, so the smallest attainable $p$-value is $1/|\mathcal{C}|!$: $1/120$ for emotion, and $1/2$ for gender, which therefore cannot be tested this way. Table~\ref{tab:appendix_exist_p} reports the resulting $p$-values, and Table~\ref{tab:appendix_ratio} the mean matched-pair and mismatched-pair overlaps behind them.

\begin{table*}[t]
  \centering
  \small
  \begin{tabular}{lccccccccc}
    \toprule
    & \multicolumn{3}{c}{\textbf{Qwen2.5-Omni}} & \multicolumn{3}{c}{\textbf{DeSTA2.5-Audio}} & \multicolumn{3}{c}{\textbf{Audio Flamingo 3}} \\
    \cmidrule(lr){2-4} \cmidrule(lr){5-7} \cmidrule(lr){8-10}
    \textbf{Task} & Matched & Mismatched & Ratio & Matched & Mismatched & Ratio & Matched & Mismatched & Ratio \\
    \midrule
    animal   & 0.245 & 0.093 & 2.62 & 0.094 & 0.075 & 1.26 & 0.264 & 0.100 & 2.63 \\
    emotion  & 0.174 & 0.113 & 1.54 & 0.121 & 0.074 & 1.65 & 0.202 & 0.115 & 1.75 \\
    gender   & 0.144 & 0.070 & 2.07 & 0.127 & 0.046 & 2.77 & 0.098 & 0.072 & 1.35 \\
    language & 0.177 & 0.089 & 1.99 & 0.125 & 0.087 & 1.43 & 0.216 & 0.090 & 2.40 \\
    \bottomrule
  \end{tabular}
  \caption{Mean Jaccard overlap of matched (diagonal) and mismatched (off-diagonal) audio-text label pairs in the neuron-overlap matrix of \S\ref{sec:exp1}, and their ratio.}
  \label{tab:appendix_ratio}
\end{table*}

\begin{table*}[t]
  \centering
  \small
  \begin{tabular}{llrcccc}
    \toprule
    \textbf{Model} & \textbf{Task} & \textbf{$k$} & \textbf{Jaccard} & \textbf{Random baseline} & \textbf{Max.\ $p$ (exist.)} & \textbf{Perm.\ $p$ (distinct.)} \\
    \midrule
    \multirow{4}{*}{Qwen2.5-Omni}      & animal   & 200 & 0.245 & 0.000185 & $1.4\times10^{-139}$ & $2.8\times10^{-6}\,\dagger$ \\
                                       & emotion  & 100 & 0.174 & 0.000096 & $2.2\times10^{-55}$ & 0.042 \\
                                       & gender   & 200 & 0.144 & 0.000185 & $4.0\times10^{-100}$ & -- \\
                                       & language & 100 & 0.177 & 0.000096 & $2.2\times10^{-55}$ & $2.5\times10^{-5}\,\dagger$ \\
    \midrule
    \multirow{4}{*}{DeSTA2.5-Audio}    & animal   & 100 & 0.094 & 0.000111 & $6.0\times10^{-30}$ & 0.0025 \\
                                       & emotion  & 100 & 0.121 & 0.000111 & $8.8\times10^{-45}$ & $0.0083\,\dagger$ \\
                                       & gender   & 200 & 0.127 & 0.000214 & $1.1\times10^{-94}$ & -- \\
                                       & language & 100 & 0.125 & 0.000111 & $9.2\times10^{-36}$ & 0.010 \\
    \midrule
    \multirow{4}{*}{Audio Flamingo 3}  & animal   & 100 & 0.264 & 0.000096 & $4.1\times10^{-89}$ & $2.8\times10^{-6}\,\dagger$ \\
                                       & emotion  & 100 & 0.202 & 0.000096 & $1.6\times10^{-68}$ & $0.0083\,\dagger$ \\
                                       & gender   & 200 & 0.098 & 0.000185 & $2.3\times10^{-74}$ & -- \\
                                       & language & 100 & 0.216 & 0.000096 & $1.6\times10^{-68}$ & $2.5\times10^{-5}\,\dagger$ \\
    \bottomrule
  \end{tabular}
  \caption{Existence of cross-modal concept neurons (\S\ref{sec:method}): tuned neuron-set size $k$, mean per-label Jaccard overlap on the held-out test split, the corresponding random-overlap baseline, and the largest per-label hypergeometric $p$-value among the task's labels (existence; Appendix~\ref{sec:appendix_hypergeom}), and the exact label-permutation $p$-value of the diagonal-vs-off-diagonal Jaccard gap (distinctiveness; Appendix~\ref{sec:appendix_perm}). $\dagger$: the smallest attainable $p$-value, $1/|\mathcal{C}|!$ for a task with $|\mathcal{C}|$ labels. All testable combinations are significant at $p<0.05$. Gender (two labels) cannot be tested by label permutation, since its smallest attainable $p$-value is $0.5$.}
  \label{tab:appendix_exist_p}
\end{table*}

\subsection{Random-Set Tests for Steering}\label{sec:appendix_suff}

For each label we draw $M=5$ random neuron sets, each of $N=|\cset(c_t)|$ neurons sampled uniformly without replacement from all FFN neurons of all layers excluding $\cset(c_t)$, with a seed determined by the model, task, label, and set index. Each set is shared across all carriers of the label, and the cross-modal concept neuron set and all random sets are evaluated on the same carriers, samples, and $\lambda$ values. Each set $S$ is summarized by $T(S)$, its target rate averaged over carriers and $\lambda$. \emph{Prediction-interval test} (per label). We transform $x=\mathrm{logit}(\mathrm{clip}(T,\epsilon,1-\epsilon))$ with $\epsilon=0.5/n$ for $n$ evaluated samples, and compute $t=(x_{\cset}-\bar x_R)/(s_R\sqrt{1+1/M})$, where $\bar x_R$ and $s_R$ are the mean and standard deviation of the random sets' $x$; the one-sided $p$-value is the upper tail of a $t$ distribution with $M-1$ degrees of freedom, i.e.\ a prediction-interval test of whether the cross-modal concept neuron set looks like one more random draw~\citep{hahn1991statistical,crawford1998comparing}. $p$-values are Benjamini--Hochberg corrected~\citep{benjamini1995controlling} across all 72 labels; without correction, 43 labels are significant at $p<0.05$, and the only one that loses significance is Qwen2.5-Omni's English ($p=0.035$, $q=0.058$). No label had identical scores across its random sets, for which the test would be undefined. \emph{Rank test} (across labels; \citealp{friedman1937use}). We rank the cross-modal concept neuron set among the $M+1$ sets by $T$, breaking ties at random with a fixed seed. Under the null each rank is uniform on $\{1,\dots,M+1\}$, and the one-sided $p$-value is the probability that the sum of independent such ranks, one per label of the model, is at most the observed sum, computed exactly by convolution.

\subsection{Bootstrap Stability of Necessity Windows}\label{sec:appendix_bootstrap}
Table~\ref{tab:exp3_necessity} decides selectivity from one evaluation of the test split. To quantify how stable this decision is, we resample, with replacement, the target task's test samples and, independently, each control task's test samples, 2{,}000 times~\citep{efron1993introduction}. Each resample is reused at every $N$ and for the baseline, as a single evaluation set is in the original sweep, and we recompute both decline onsets with the rule of \S\ref{sec:exp3}. A combination counts as selective in a resample when the target onset precedes the control onset. Table~\ref{tab:appendix_onset_bootstrap} reports the results: the observed onsets reproduce Table~\ref{tab:exp3_necessity}, 9 of the 10 selective combinations remain selective in at least $90\%$ of resamples, Audio Flamingo 3's gender window is selective in only $39\%$, and DeSTA2.5-Audio's animal and language tasks remain non-selective in most resamples.

\begin{table*}[t]
  \centering
  \small
  \begin{tabular}{llccccc}
    \toprule
    \textbf{Task} & \textbf{Model} & \textbf{Target onset} & \textbf{95\% interval} & \textbf{Control onset} & \textbf{95\% interval} & \textbf{P(selective)} \\
    \midrule
    \multirow{3}{*}{animal} & Qwen2.5-Omni     & 1{,}745 & 932--3{,}406 & 36{,}014 & 36{,}014--36{,}014 & 1.00 \\
                           & DeSTA2.5-Audio   & 5{,}973 & 2{,}693--40{,}280 & 5{,}973 & 2{,}693--15{,}110 & 0.25 \\
                           & Audio Flamingo 3 & 984 & 984--1{,}804 & 123{,}275 & 63{,}872--123{,}275 & 1.00 \\
    \midrule
    \multirow{3}{*}{emotion} & Qwen2.5-Omni     & 1{,}618 & 133--5{,}996 & 71{,}812 & 71{,}812--189{,}409 & 1.00 \\
                           & DeSTA2.5-Audio   & 10{,}336 & 4{,}395--26{,}763 & 26{,}763 & 26{,}763--26{,}763 & 0.90 \\
                           & Audio Flamingo 3 & 5{,}862 & 444--69{,}437 & 180{,}203 & 69{,}437--180{,}203 & 1.00 \\
    \midrule
    \multirow{3}{*}{gender} & Qwen2.5-Omni     & 1{,}255 & 669--1{,}255 & 264{,}115 & 264{,}115--264{,}115 & 1.00 \\
                           & DeSTA2.5-Audio   & 317 & 199--317 & 93{,}948 & 93{,}948--93{,}948 & 1.00 \\
                           & Audio Flamingo 3 & 74{,}073 & 25{,}037--251{,}664 & 251{,}664 & 74{,}073--251{,}664 & 0.39 \\
    \midrule
    \multirow{3}{*}{language} & Qwen2.5-Omni     & 3{,}817 & 1{,}936--3{,}817 & 107{,}461 & 107{,}461--107{,}461 & 1.00 \\
                           & DeSTA2.5-Audio   & 5{,}430 & 5{,}430--33{,}242 & 5{,}430 & 5{,}430--12{,}879 & 0.03 \\
                           & Audio Flamingo 3 & 19{,}538 & 19{,}538--19{,}538 & 102{,}782 & 48{,}173--102{,}782 & 1.00 \\
    \bottomrule
  \end{tabular}
  \caption{Bootstrap stability of the necessity windows of Table~\ref{tab:exp3_necessity} (2{,}000 resamples of the test samples). Onsets are numbers of deactivated neurons $N$; P(selective) is the fraction of resamples in which the target task's decline onset precedes the control tasks'.}
  \label{tab:appendix_onset_bootstrap}
\end{table*}

\section{Additional Experimental Setup Details}

This appendix collects experimental-setup details referenced from \S\ref{sec:setup} and \S\ref{sec:results} but not spelled out in the main text.

\subsection{Neuron-Set Size for Steering}\label{sec:appendix_ksuff}
For each label, we repeat the deactivation sweep of \S\ref{sec:exp3} at the label level: we deactivate only that label's $\cset(c)$ and compare accuracy on that label (target) with the mean accuracy on the task's other labels (control), locating both decline onsets as in \S\ref{sec:exp3}. $k$ (and hence $N$) is the value between the two onsets that maximizes the target's relative accuracy drop minus the control's; labels without such a window use a fixed size matched to the task's strongest-window label.

\subsection{Synthetic Carriers}\label{sec:appendix_synthcarrier}
Silence and noise are not labels of any task, so for these two carriers we compute $v^{\text{text}}(c_r)$ as for a real label, but from five sentences describing silence (e.g.\ ``This audio contains complete silence.'') or noise (e.g.\ ``This audio contains only random noise.'').

\subsection{Audio-Only Reference Control}
The audio-only reference control of \S\ref{sec:exp3} gives each label an equal quota of $N/|\mathcal{C}|$ neurons from its own audio top-$k$ first, then adds, in round-robin order across labels, each label's next-ranked audio neuron not yet selected until the union reaches size $N$, since the equal-quota union alone can fall short of $N$.

\subsection{Number of Deactivated Neurons}\label{sec:appendix_nablated}
Deactivation sweeps the per-label cutoff $k$ of \S\ref{sec:method}, but the deactivated set $\cset=\bigcup_{c}\cset(c)$ is a union of per-label intersections, so its size $N=|\cset|$ differs from $k$: the intersections shrink it and the union over labels enlarges it. We therefore report deactivation results against $N$, to which every random-neuron and audio-only control is matched. $N$ increases strictly with $k$ for every model and task, so the decline onsets and the selectivity decisions do not depend on which of the two is used. Qwen2.5-Omni's target onsets correspond to $0.24$--$0.72\%$ of its $530{,}432$ FFN neurons (28 layers $\times$ 18{,}944).

\subsection{Synonym Control}\label{sec:appendix_synonym}
Because each label's text prompts contain the label word itself (e.g.\ ``The speaker feels angry.''), a cross-modal concept neuron set could in principle reflect the literal label word rather than the attribute. To test this, we re-extract Qwen2.5-Omni's text activations for the emotion task with the same five templates but the label word replaced by a synonym sharing no token with it (angry $\to$ furious, happy $\to$ joyful, sad $\to$ sorrowful, fear $\to$ terror, disgust $\to$ revulsion), re-identify cross-modal concept neurons at $k \in \{1{,}000, 2{,}000, 4{,}000\}$, and repeat the task-level deactivation of \S\ref{sec:exp3}. The synonym-based sets overlap substantially with the original ones (mean per-label Jaccard $0.52$--$0.53$, against $0.04$--$0.05$ between mismatched labels), and deactivating them is nearly as damaging to emotion recognition as deactivating the original sets (relative target-accuracy drop $8.1$/$11.2$/$16.1\%$ vs.\ $8.7$/$13.0$/$18.6\%$ at the three values of $k$, against $2.2$--$3.9\%$ for matched-size random neurons), with similar control-task drops ($5.1$--$5.8\%$ vs.\ $3.7$--$7.4\%$). At least for this model and task, the neurons thus track the attribute rather than the specific label word.

\subsection{Audio-Position Control}\label{sec:appendix_audiopos}
Both interventions are applied at every token position, so part of their effect could come from disrupting how the model reads the question rather than how it processes the audio. We therefore repeat both on Qwen2.5-Omni's emotion task, the same setting as the synonym control, restricting the intervention to each sample's audio-token positions. Deactivation remains selective: the target task's decline onset moves from $N=1{,}618$ to $12{,}804$ and the control tasks' from $71{,}812$ to $189{,}409$, the control tasks are nearly unaffected until $N=71{,}812$, and at $N=1{,}618$ and $12{,}804$ target accuracy is below all five random-neuron control trials, whose neurons are likewise deactivated only at audio-token positions. For steering, the cross-modal concept neuron set outranks all five random sets for all five labels (rank test $p=1.3\times10^{-4}$) and is significant for four of them under the prediction-interval test; its lead over the random sets in average target rate is $54$--$82\%$ as large as under whole-sequence steering. This holds even for silent input, which carries no acoustic content: at $\lambda=5$ the cross-modal concept neuron set's target rate exceeds every random set's for all five labels, reaching $100\%$ for ``happy'' against at most $30\%$ for the random sets. Restricting the interventions to audio positions thus weakens both effects, but leaves their selectivity, their advantage over random neurons, and hence our conclusions unchanged.

\subsection{Layer-Matched Random Control}\label{sec:appendix_layermatched}
Cross-modal concept neurons concentrate in later layers ($48$--$73\%$ of each Qwen2.5-Omni emotion set lies in the last quarter of the layers), whereas the random sets of \S\ref{sec:exp4} are drawn uniformly across layers, so the cross-modal concept neuron set's advantage could in principle reflect only where its neurons sit. We therefore repeat the steering test on Qwen2.5-Omni's emotion task, the same setting as the other controls, with five random sets that match the cross-modal concept neuron set's number of neurons in every layer. The cross-modal concept neuron set outranks all five layer-matched random sets for all five labels (rank test $p=1.3\times10^{-4}$) and is significant for four of them under the prediction-interval test; its lead over the random sets in average target rate ($0.17$--$0.35$) is essentially unchanged from the uniform random sets ($0.17$--$0.38$). Its advantage therefore does not come from its layer distribution.

\subsection{DeSTA2.5-Audio Transcript Input}\label{sec:appendix_desta_asr}
To check what DeSTA2.5-Audio's transcript input (\S\ref{sec:setup}) reveals, we ran its own transcription pipeline (voice-activity detection followed by Whisper-large-v3 decoding) on all 500 clips of each task. For spoken language, every clip receives a transcript, written in the spoken language: judging the transcript's language by its script (Chinese, Japanese, Korean) or by common function words (European languages), $83\%$ match the label ($100\%$ for Chinese and Korean, $62$--$89\%$ for the European languages and Japanese, where short transcripts often contain no decisive word), a lower bound on how often the answer is readable from text. For animal sounds, $94\%$ of clips are rejected by the voice-activity detector and receive an empty transcript, and only about $1\%$ of transcripts name the correct animal (e.g.\ ``Woof!''). Emotion clips come from CREMA-D~\citep{cao2014cremad}, whose fixed sentences are shared across emotions ($57\%$ of transcripts), and MELD~\citep{poria2019meld}, TV-show dialogue whose content can occasionally hint at the emotion; gender transcripts contain only the spoken sentence. The transcript thus offers a systematic text route to the answer only for spoken language.

\subsection{Answer Parsing}\label{sec:appendix_parsing}
Models are prompted to answer with a single option letter. We take the first character of the stripped, lowercased response as the predicted letter, and additionally accept a leading ``(x)'' when it is the only option-shaped string in the response, since Audio Flamingo 3 frequently answers in the form ``(a) sheep''. A response that yields no letter in $\{a,b,c,d\}$ counts as a parse failure and is scored as incorrect (\S\ref{sec:exp3}) or as not predicting the target label (\S\ref{sec:exp4}), identically in every condition. Parse failures are rare. None of the three models produces any at baseline, and at each task's target decline onset (Table~\ref{tab:exp3_necessity}) cross-modal-neuron deactivation produces none except in Audio Flamingo 3's gender task ($2.0\%$ of responses, accounting for $0.8$ of its $6.2$ percentage-point accuracy drop), so the necessity results reflect changed answers rather than broken outputs. Under contrastive neuron steering (\S\ref{sec:exp4}), parse failures rise at the largest coefficient, more for cross-modal concept neurons ($7.9$--$16.4\%$ of responses) than for random neurons ($0.1$--$4.4\%$). Because a parse failure never counts as predicting the target label, this can only lower the cross-modal concept neurons' target rate, biasing the comparison against rather than toward our steering results.

\subsection{Text Descriptions}\label{sec:appendix_prompts}
Each label's five paraphrased text descriptions follow a fixed template per task, substituting only the label word; the one exception is emotion, where the noun-form labels (\emph{fear}, \emph{disgust}) use a slightly different template than the adjective-form labels (\emph{happy}, \emph{sad}, \emph{angry}) for grammaticality. One representative label per task:

\begin{itemize}
  \item \textbf{Emotion} (\emph{angry}): ``The speaker feels angry.''; ``The speaker sounds angry right now.''; ``This is an angry voice.''; ``The speaker is expressing an angry tone.''; ``The overall emotion in the speaker's voice is angry.''
  \item \textbf{Emotion} (\emph{fear}, noun-form template): ``The speaker feels fear.''; ``The speaker's voice is full of fear.''; ``This voice conveys a sense of fear.''; ``The speaker is expressing fear.''; ``The overall emotion in the speaker's voice is fear.''
  \item \textbf{Animal} (\emph{dog}): ``The sound in the audio is from a dog.''; ``This audio contains the sound of a dog barking.''; ``A dog is making the sound in this audio.''; ``The animal making this sound is a dog.''; ``This is the sound of a dog.''
  \item \textbf{Gender} (\emph{female}): ``The speaker in the audio is female.''; ``This is a female voice.''; ``The voice belongs to a female speaker.''; ``The speaker is female.''; ``The gender of the speaker is female.''
  \item \textbf{Language} (\emph{English}): ``The speaker is speaking English.''; ``This audio is in the English language.''; ``The language spoken in the audio is English.''; ``The speaker is using English.''; ``English is being spoken in this audio.''
\end{itemize}
All other labels within a task use the identical template with only the label word (and, for animal, the corresponding sound verb, e.g.\ \emph{barking} $\to$ \emph{meowing} for \emph{cat}) substituted.

\section{Additional Figures}

This appendix collects figures referenced from \S\ref{sec:results} but not shown in the main text.

\subsection{Cross-Modal Neuron-Jaccard Overlap Matrices}
\label{sec:appendix_exp2matrix}

\begin{figure*}[t]
  \centering
  \includegraphics[width=0.48\textwidth]{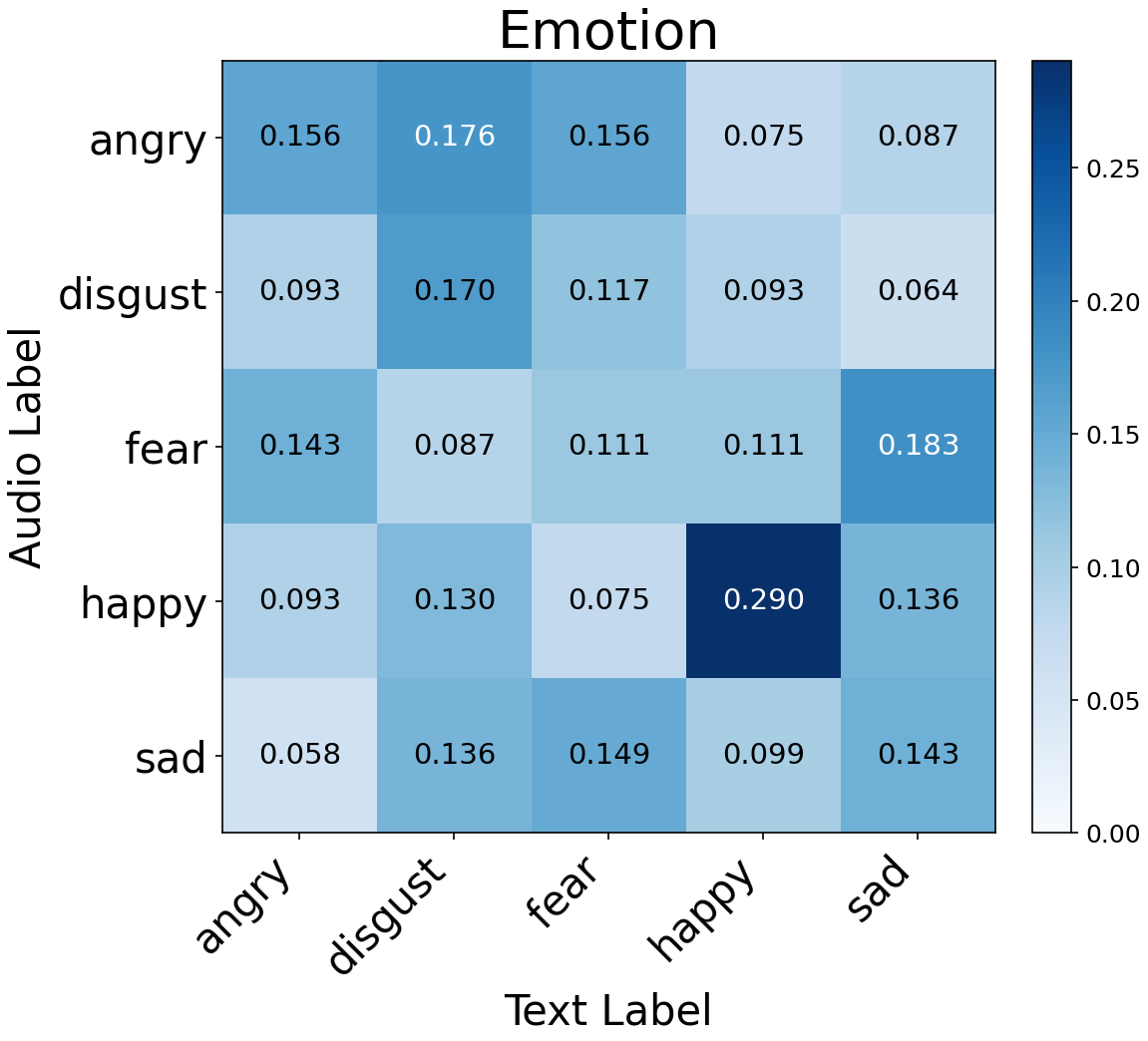}\hfill
  \includegraphics[width=0.48\textwidth]{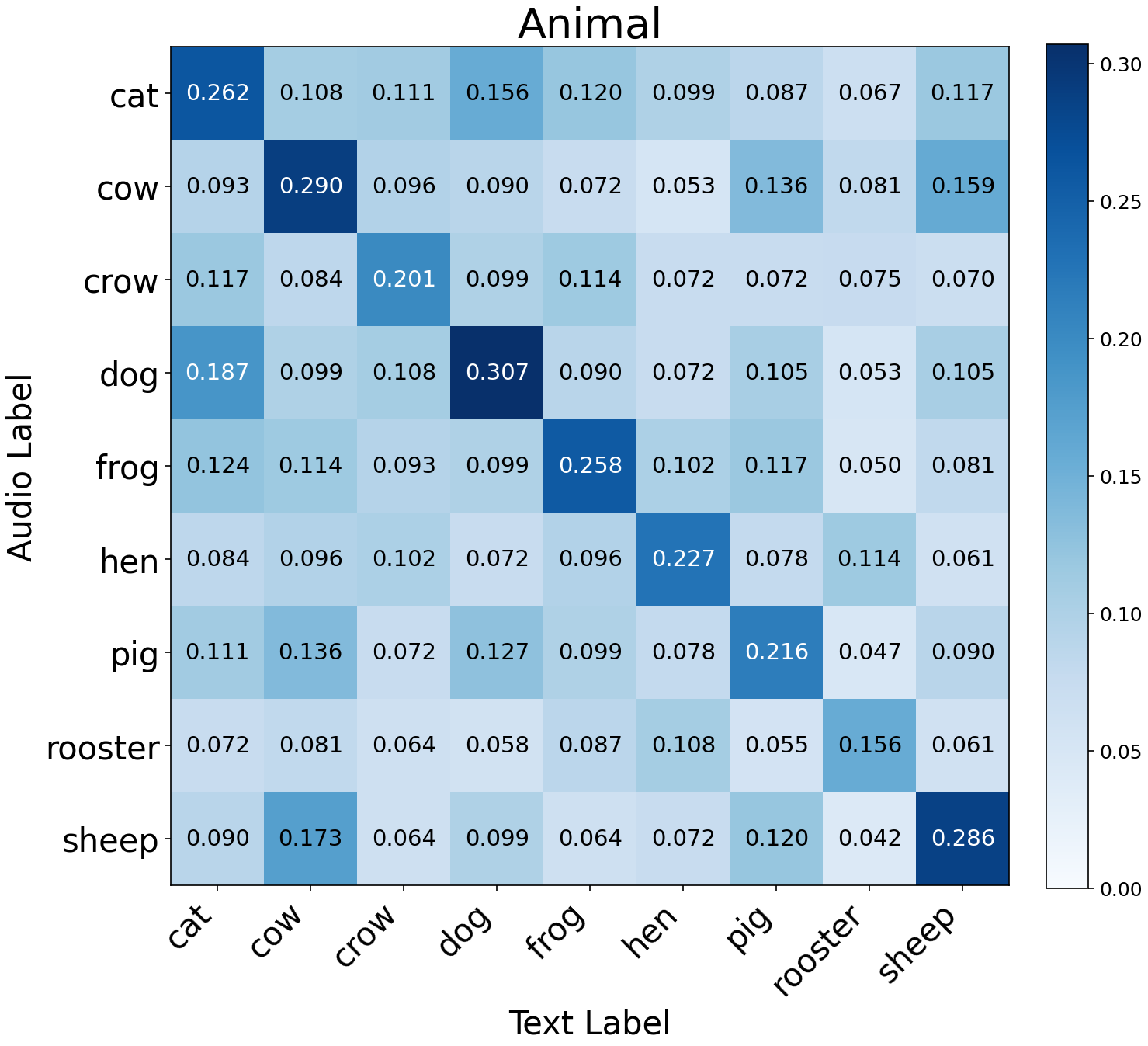}\\[2mm]
  \includegraphics[width=0.48\textwidth]{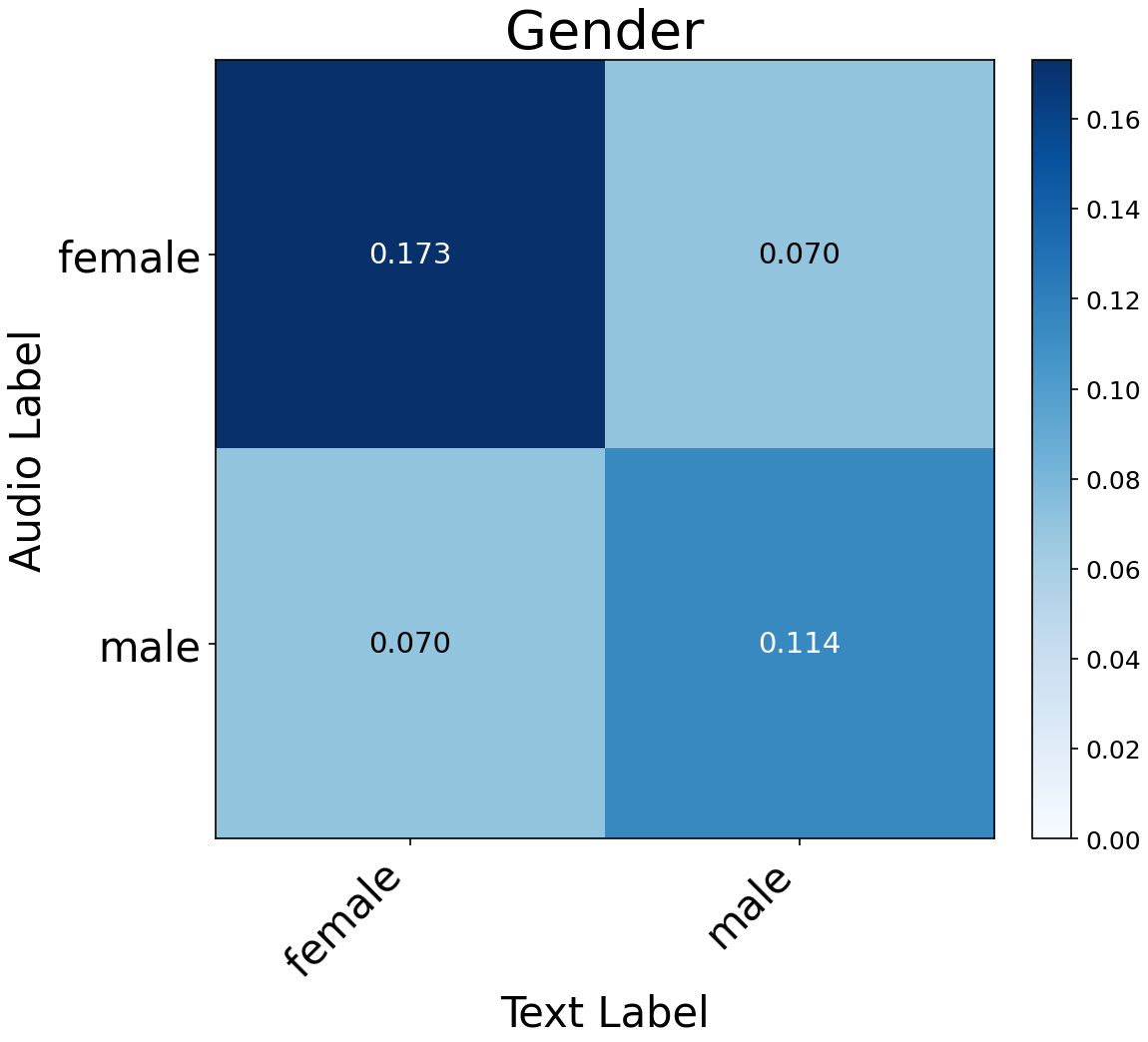}\hfill
  \includegraphics[width=0.48\textwidth]{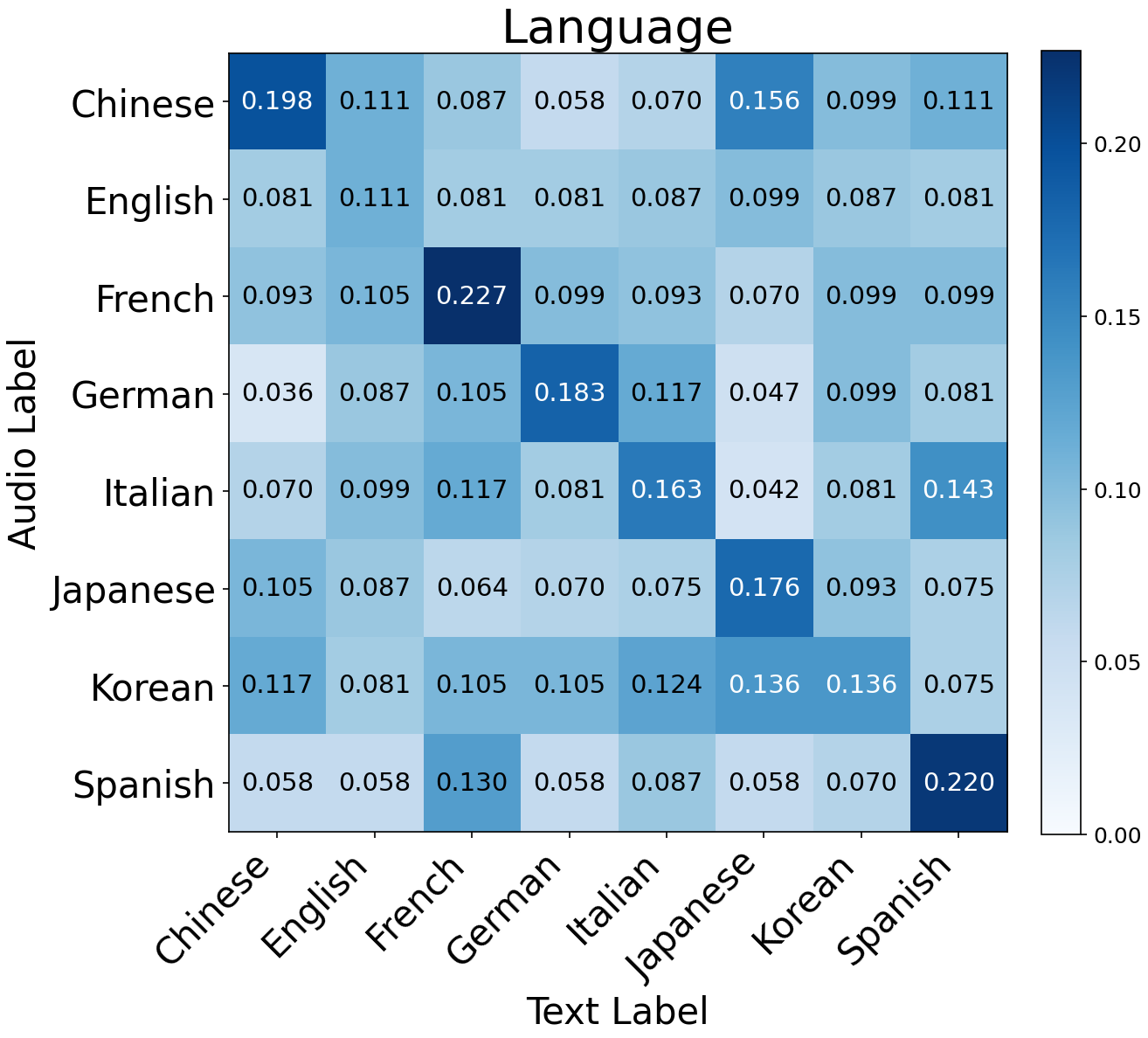}
  \caption{Neuron-Jaccard overlap matrices for Qwen2.5-Omni, by task.}
  \label{fig:exp2matrix_qwen}
\end{figure*}

\begin{figure*}[t]
  \centering
  \includegraphics[width=0.48\textwidth]{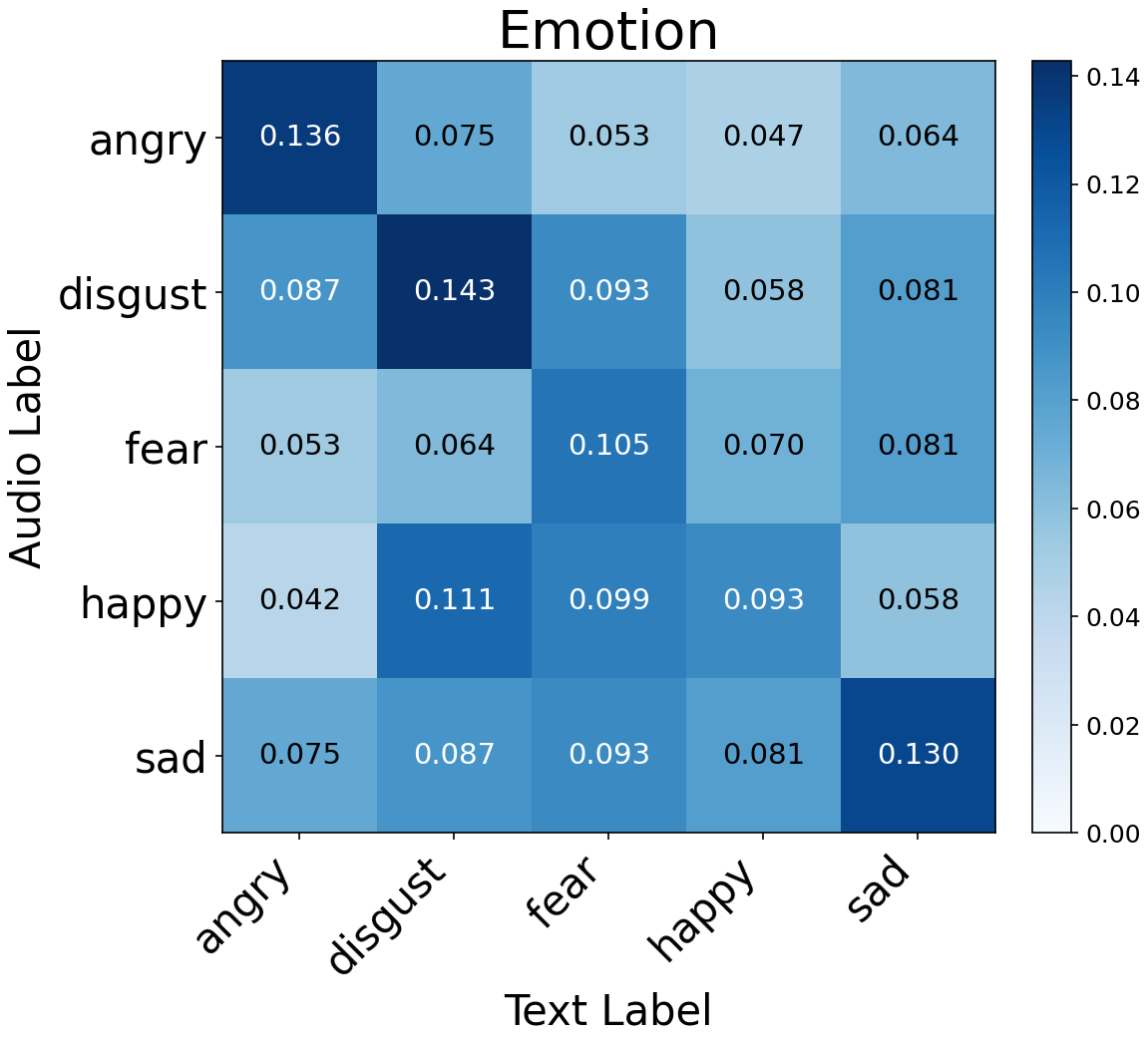}\hfill
  \includegraphics[width=0.48\textwidth]{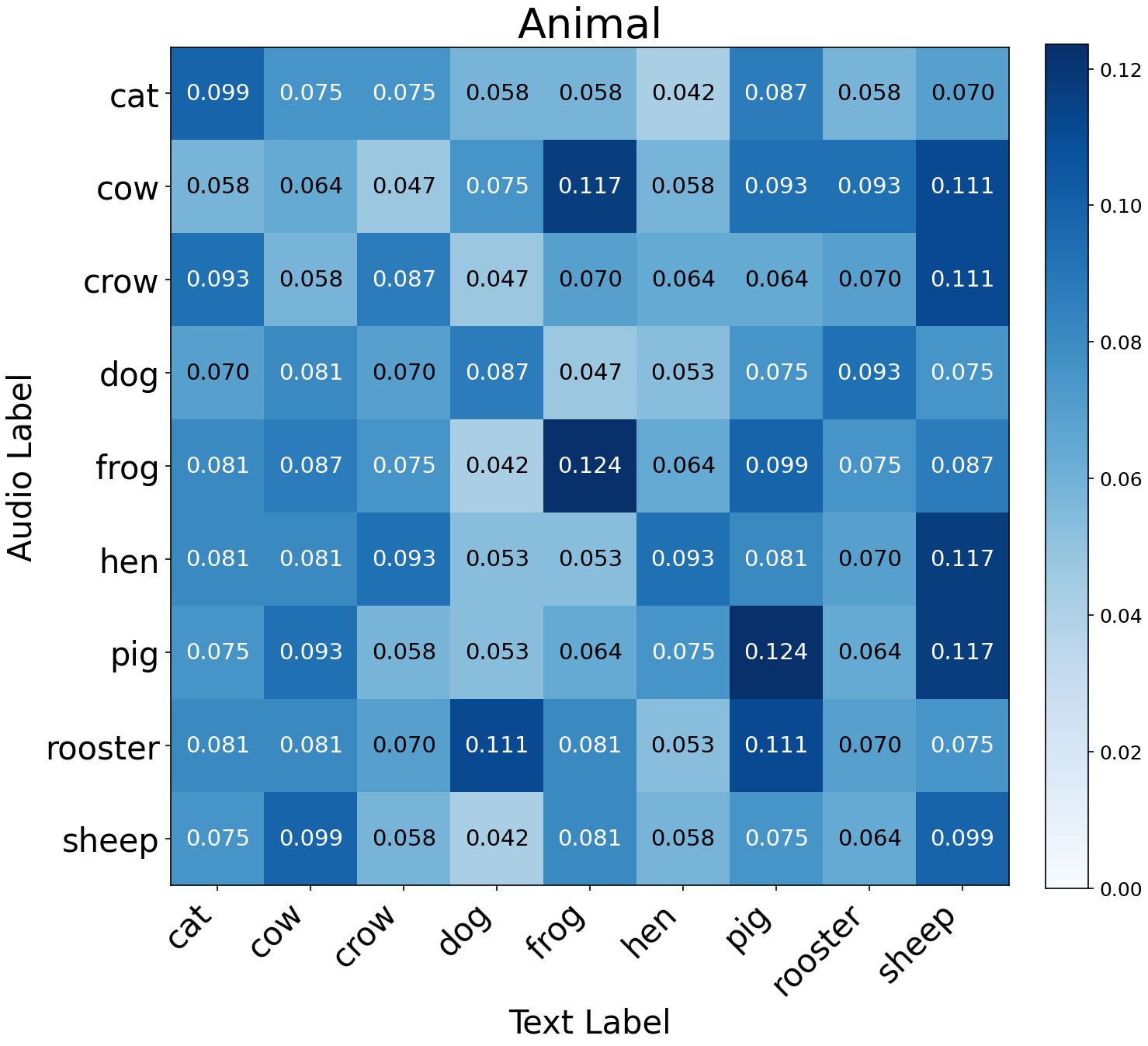}\\[2mm]
  \includegraphics[width=0.48\textwidth]{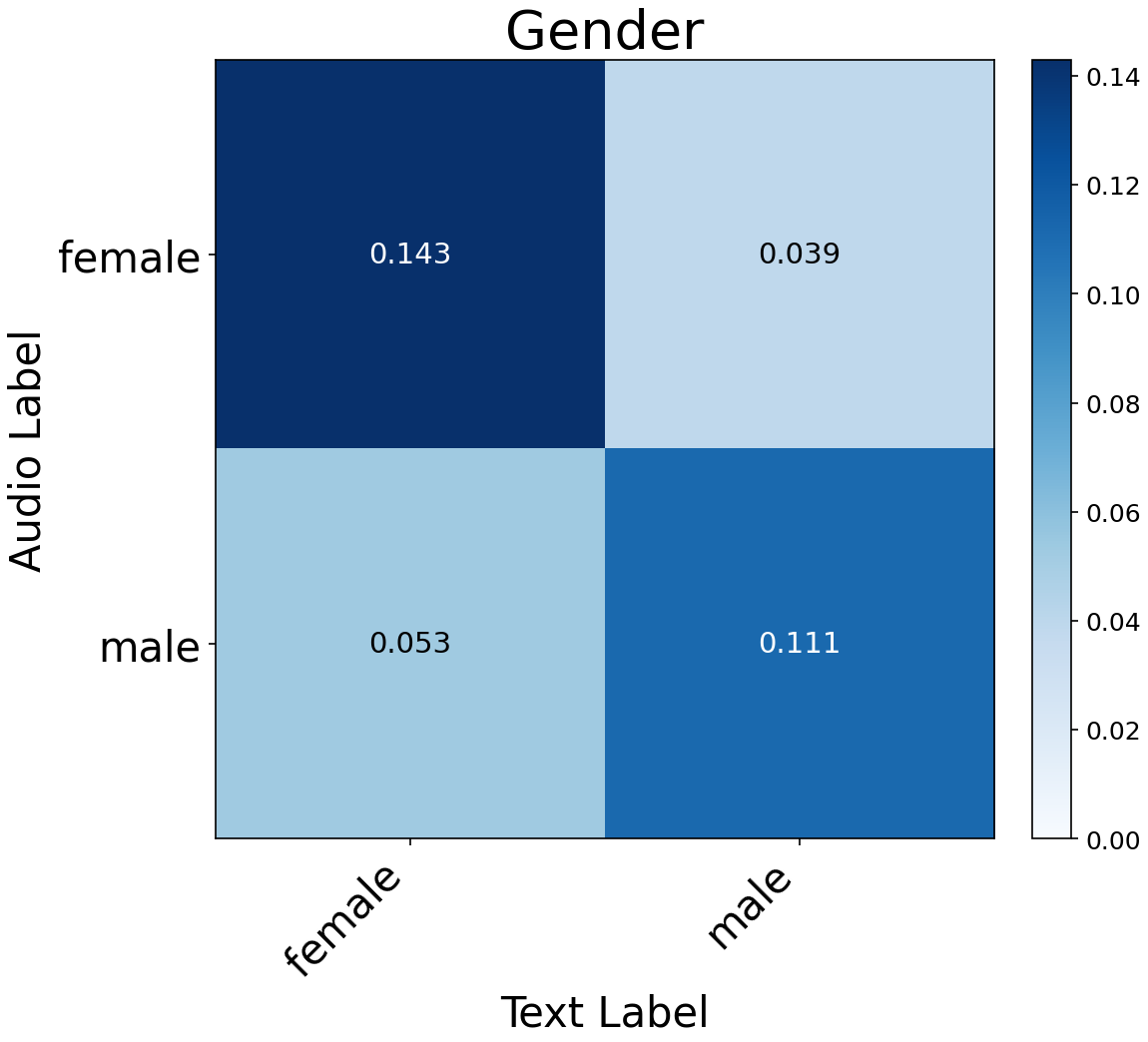}\hfill
  \includegraphics[width=0.48\textwidth]{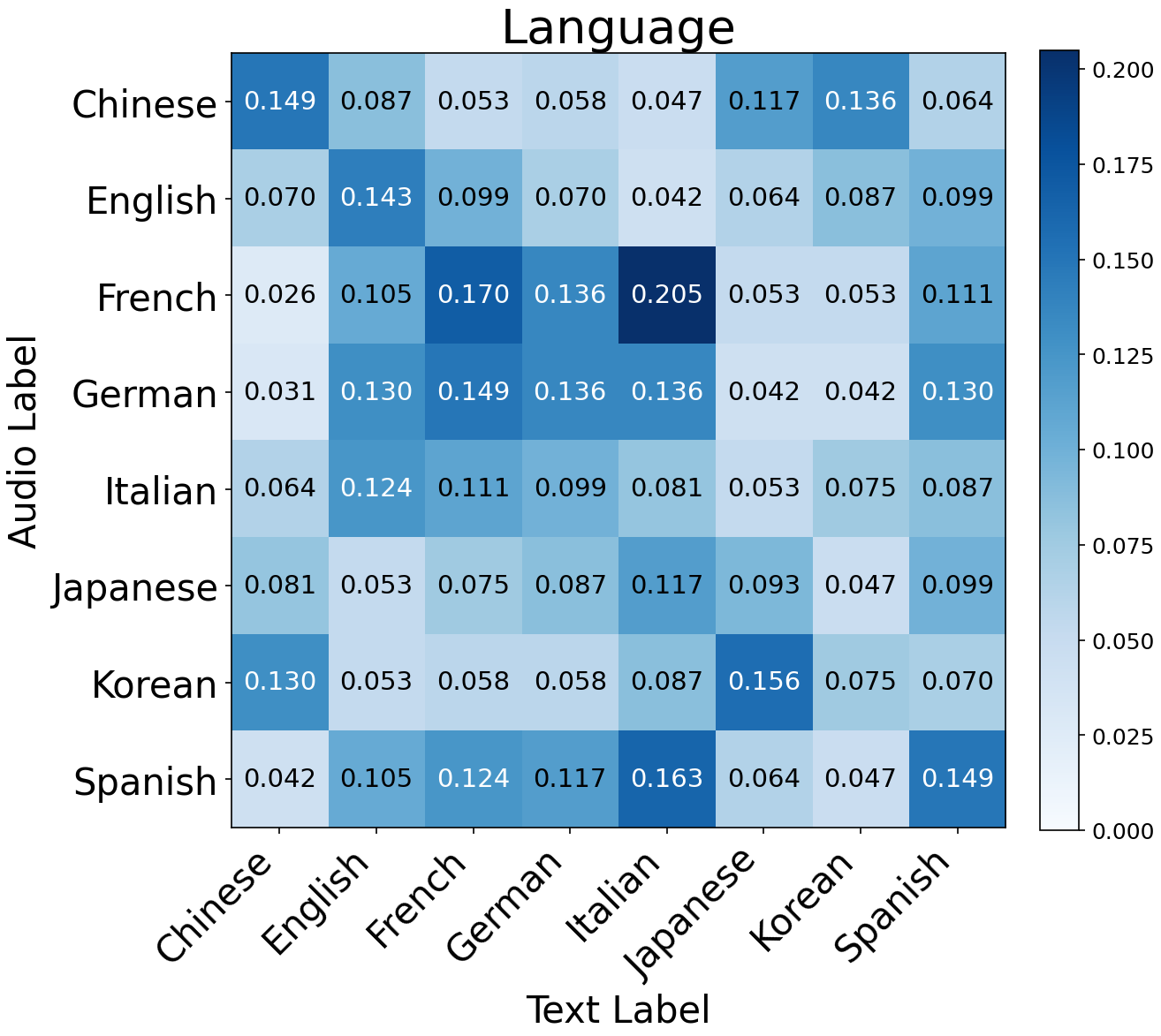}
  \caption{Neuron-Jaccard overlap matrices for DeSTA2.5-Audio, by task.}
  \label{fig:exp2matrix_desta}
\end{figure*}

\begin{figure*}[t]
  \centering
  \includegraphics[width=0.48\textwidth]{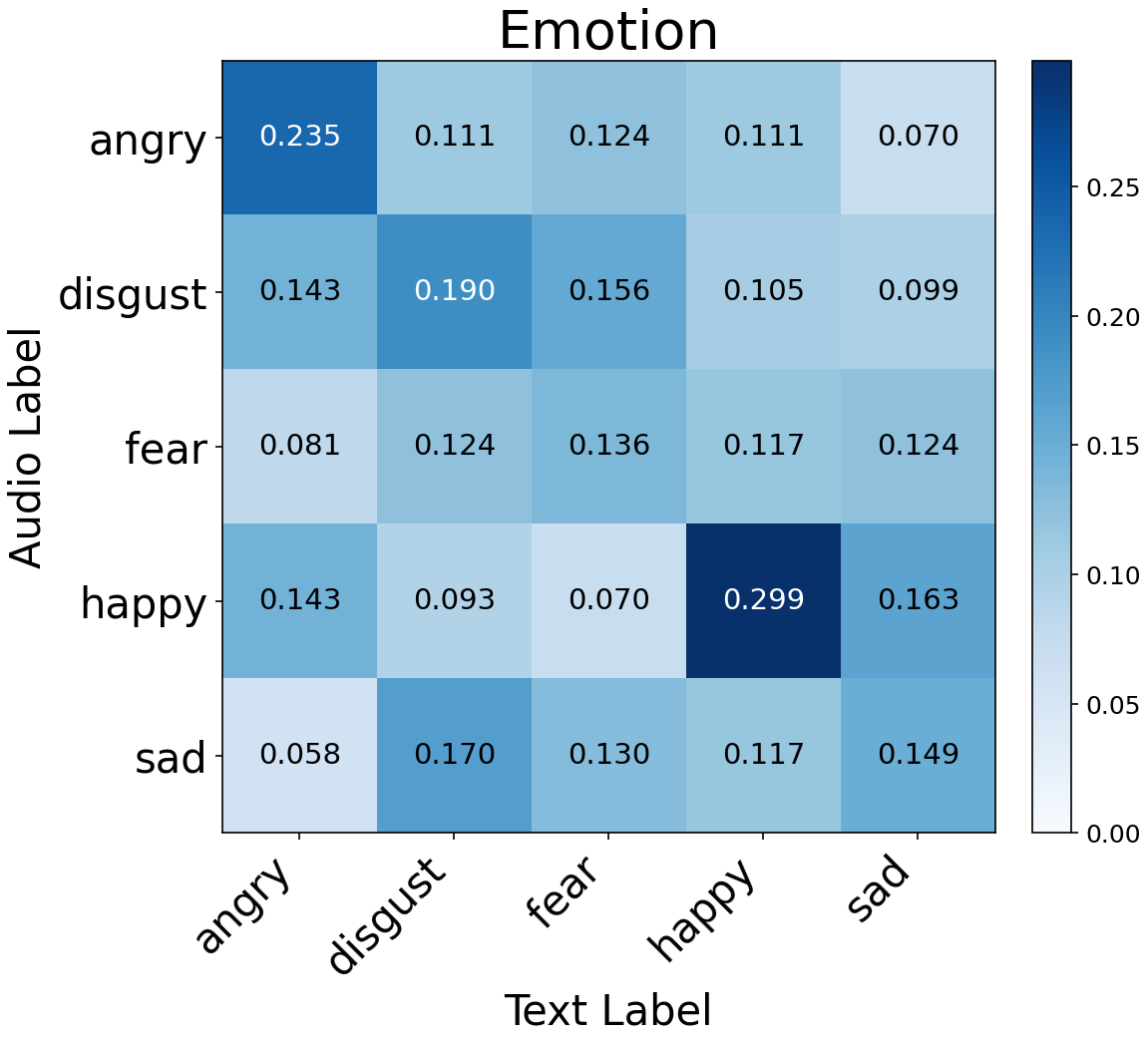}\hfill
  \includegraphics[width=0.48\textwidth]{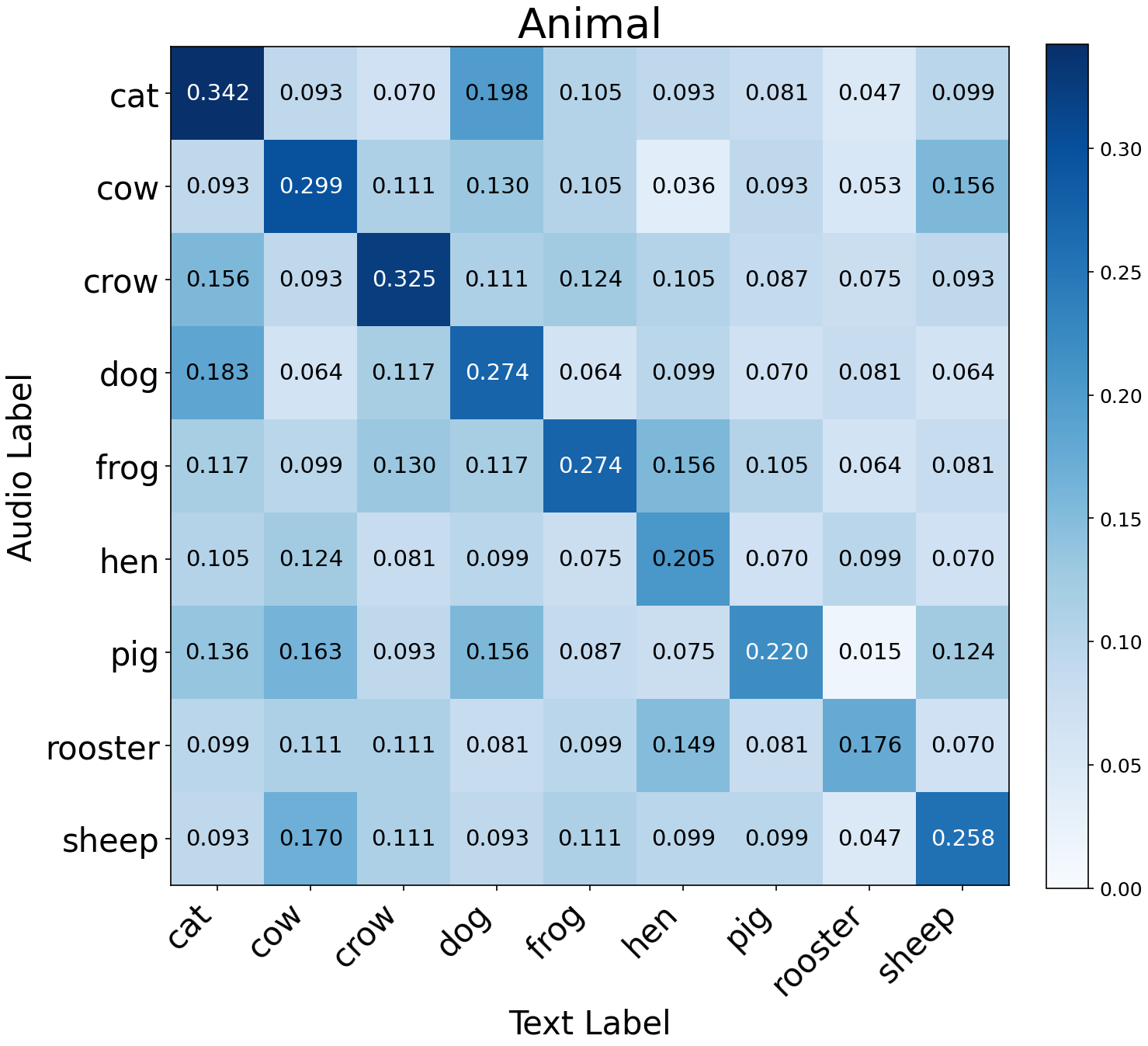}\\[2mm]
  \includegraphics[width=0.48\textwidth]{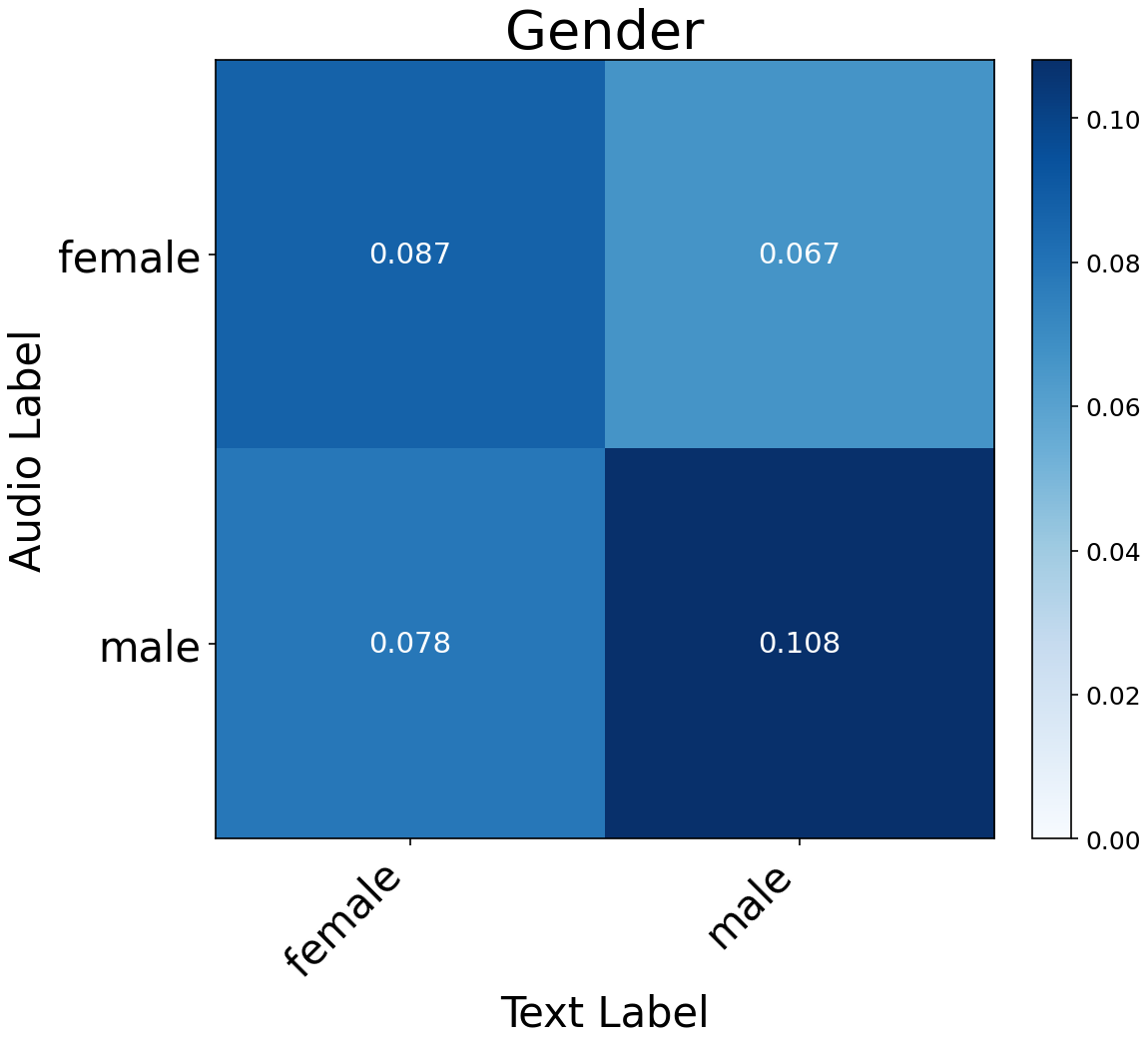}\hfill
  \includegraphics[width=0.48\textwidth]{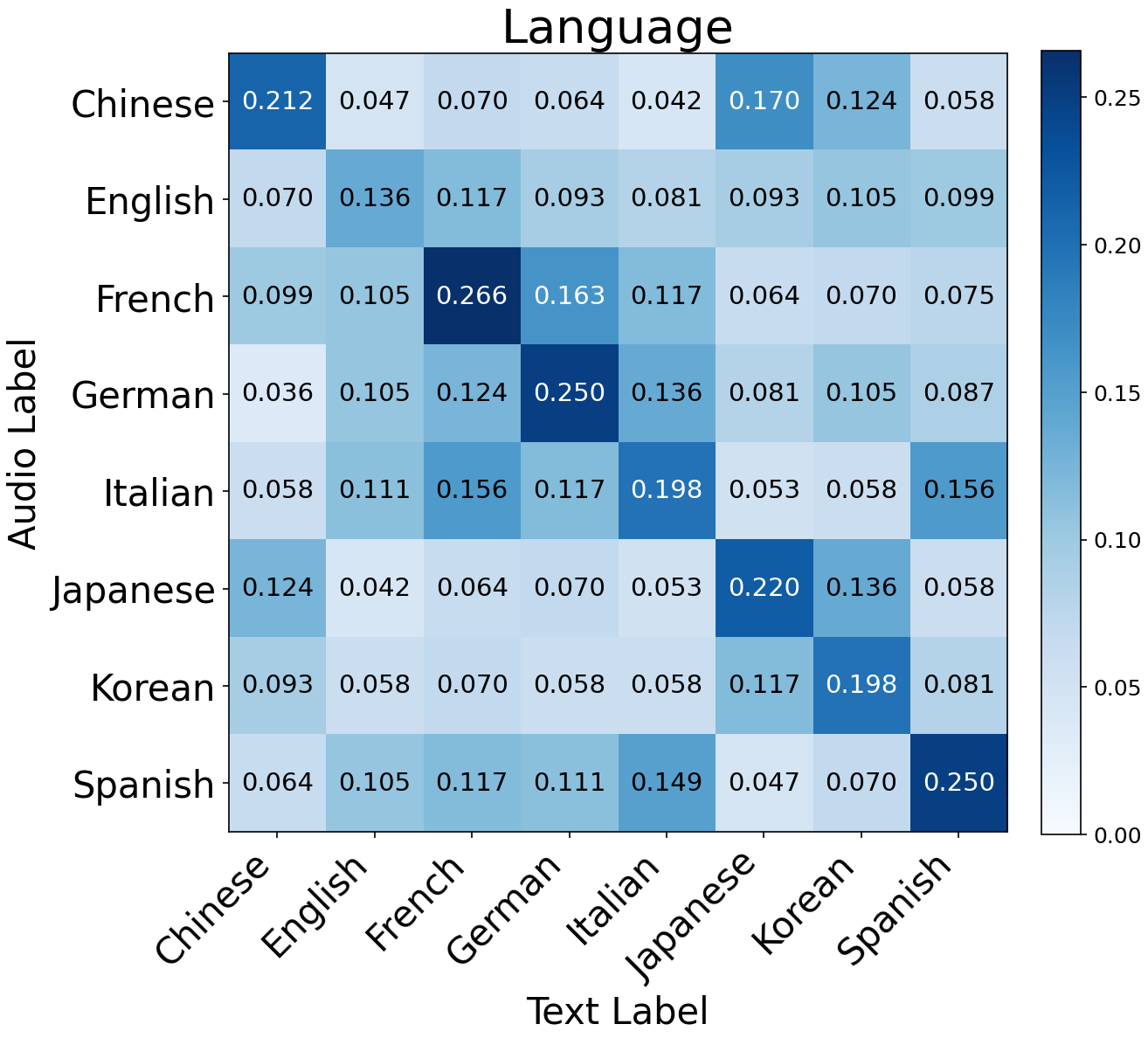}
  \caption{Neuron-Jaccard overlap matrices for Audio Flamingo 3, by task.}
  \label{fig:exp2matrix_af3}
\end{figure*}

Figures~\ref{fig:exp2matrix_qwen}--\ref{fig:exp2matrix_af3} give the full label-by-label neuron-Jaccard overlap matrix underlying the distinctiveness test in \S\ref{sec:exp1}, one figure per model, with one heatmap per task. Rows are audio labels, columns are text labels; a cell is the Jaccard overlap between that audio label's and that text label's top-$k$ neuron sets. The diagonal entries are the matched-label overlaps summarized in Table~\ref{tab:appendix_ratio} (as a diagonal-to-off-diagonal ratio), Table~\ref{tab:variance_share} (as the matched-label share), and Table~\ref{tab:exp1_existence} (as a mean per-label Jaccard overlap, for all four tasks including gender), and are visibly higher than most off-diagonal entries in every task and model, consistent with the significant diagonal-vs-off-diagonal gap reported there.

\subsection{Steering Curves: the Same Label across Models}
\label{sec:appendix_exp4}

\begin{figure*}[t]
  \centering
  \includegraphics[width=0.8\textwidth]{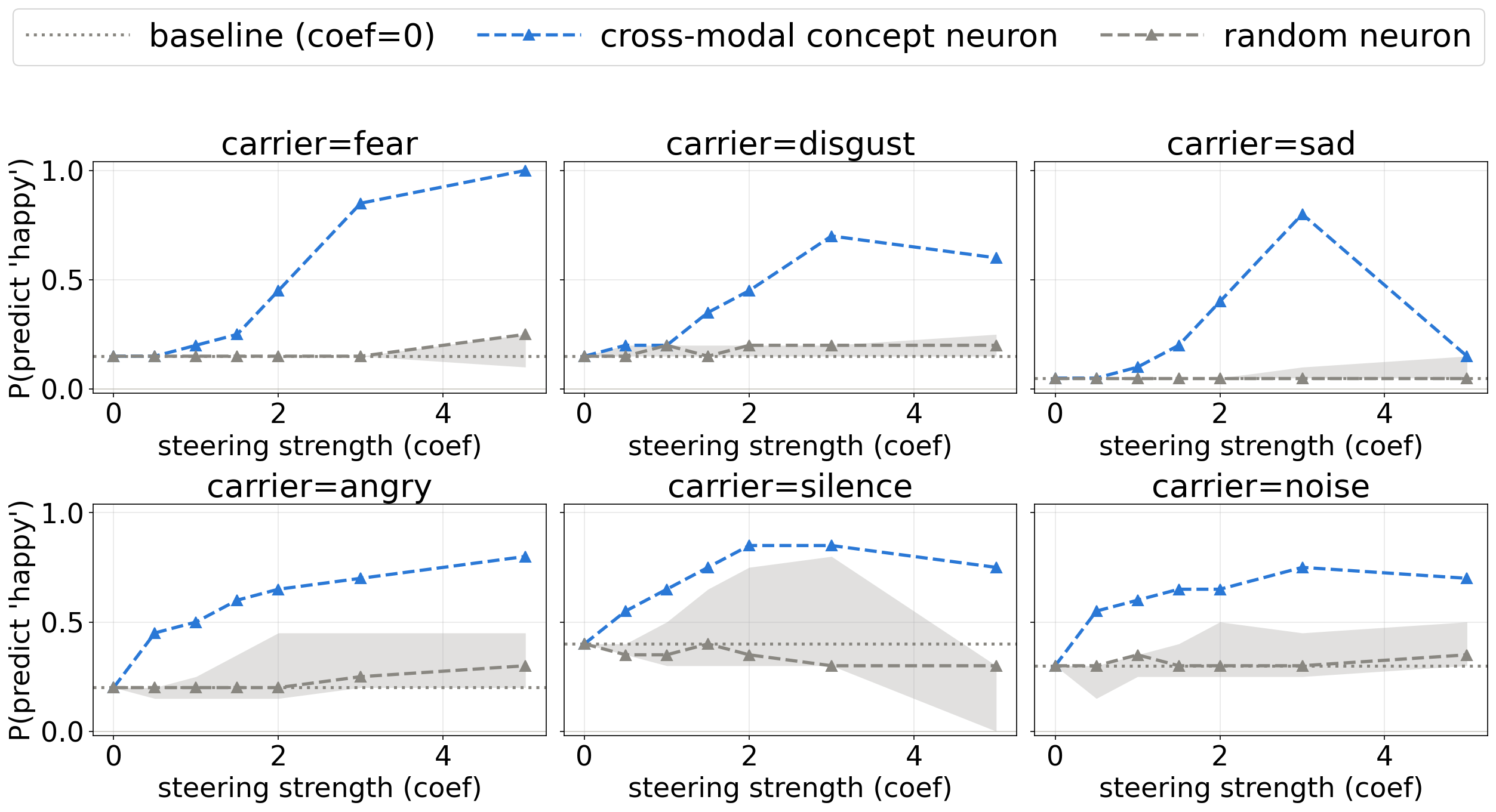}
  \caption{Contrastive neuron steering toward the emotion label ``happy'' in DeSTA2.5-Audio, one panel per carrier label. $x$-axis: steering coefficient; $y$-axis: fraction of samples whose predicted answer is ``happy''. Blue: cross-modal concept neurons; grey: median of five matched-size random neuron sets, with their min--max range shaded (the null).}
  \label{fig:exp4_desta_happy}
\end{figure*}

\begin{figure*}[t]
  \centering
  \includegraphics[width=0.8\textwidth]{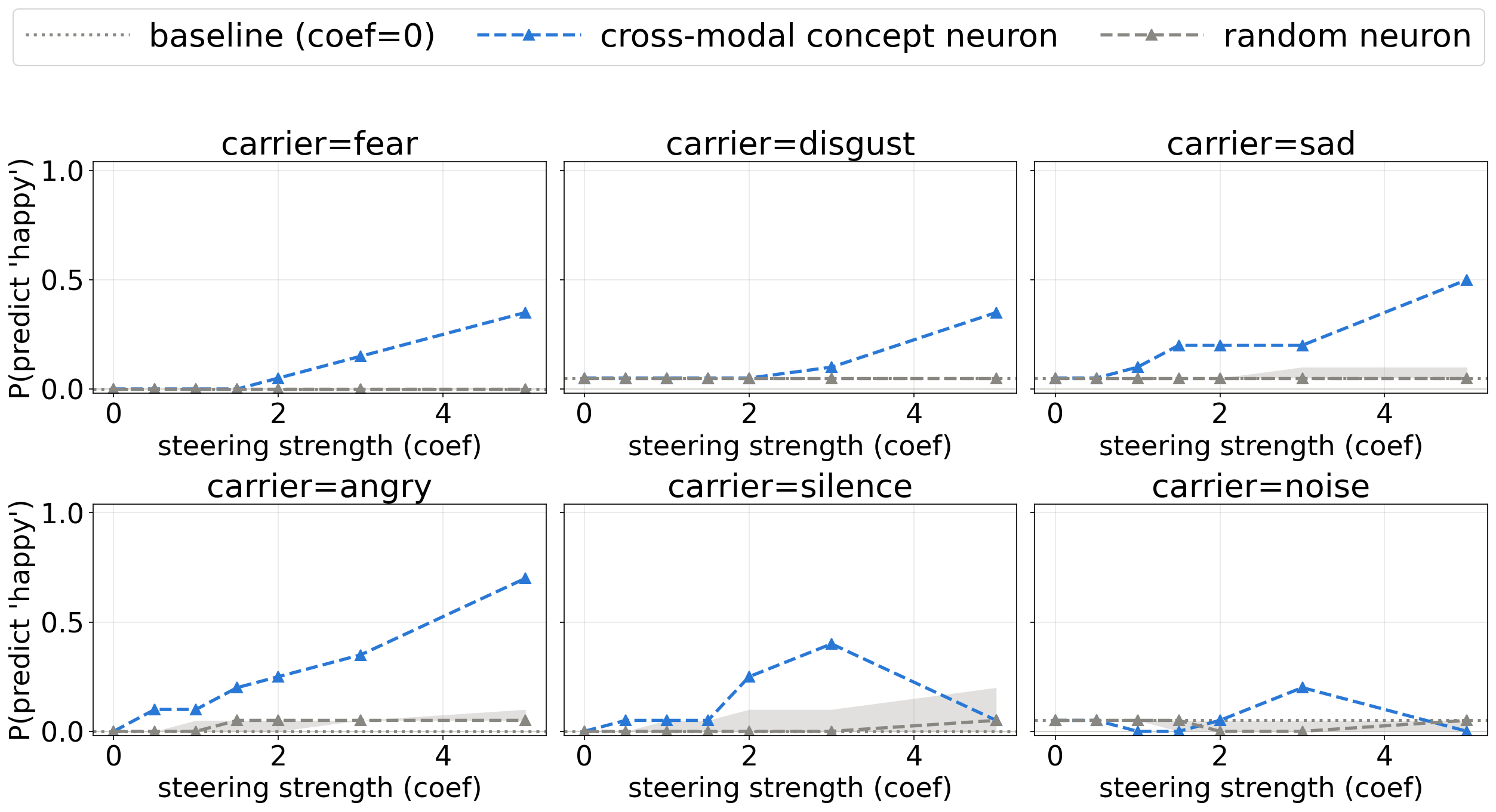}
  \caption{Contrastive neuron steering toward the emotion label ``happy'' in Audio Flamingo 3, one panel per carrier label. $x$-axis: steering coefficient; $y$-axis: fraction of samples whose predicted answer is ``happy''. Blue: cross-modal concept neurons; grey: median of five matched-size random neuron sets, with their min--max range shaded (the null).}
  \label{fig:exp4_af3_happy}
\end{figure*}

Figures~\ref{fig:exp4_desta_happy} and~\ref{fig:exp4_af3_happy} give DeSTA2.5-Audio's and Audio Flamingo 3's steering curves for the same label shown for Qwen2.5-Omni in Figure~\ref{fig:exp4_qwen_happy} (\S\ref{sec:exp4}), emotion label ``happy.'' Both models also show a positive gap between cross-modal concept neurons and random neurons on this label, though smaller than Qwen2.5-Omni's.

\subsection{Deactivation Curves for the Remaining Tasks}
\label{sec:appendix_exp3}

\begin{figure*}[t]
  \centering
  \includegraphics[width=\textwidth]{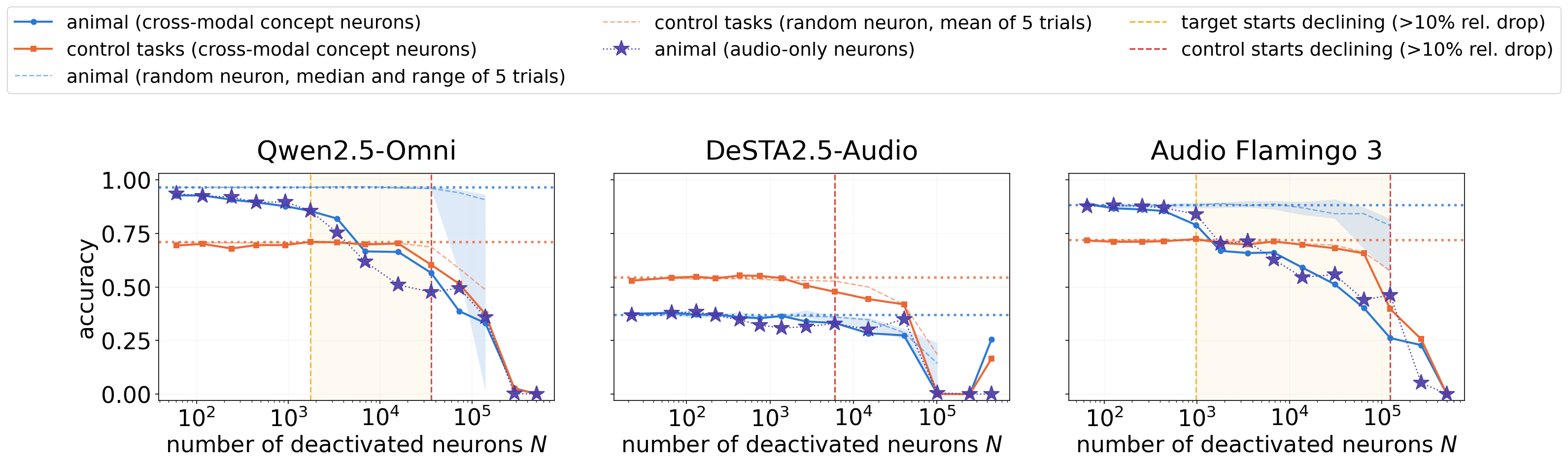}
  \caption{Task-level deactivation on the animal task, for Qwen2.5-Omni, DeSTA2.5-Audio, and Audio Flamingo 3 (left to right). Solid lines: target task accuracy (blue) and mean control-task accuracy (orange) as a function of the number of deactivated neurons $N$ (log scale). Dashed lines: matched-size random-neuron control (target task: median of the five random trials, with their min--max range shaded; control tasks: mean). Purple stars: audio-only neurons, a reference control (\S\ref{sec:exp3}). Vertical dashed lines and the shaded band mark the target's and controls' decline onsets; a band is shown only when the target's onset precedes the controls'. Unlike emotion, DeSTA2.5-Audio shows no selective window here.}
  \label{fig:exp3_animal}
\end{figure*}

\begin{figure*}[t]
  \centering
  \includegraphics[width=\textwidth]{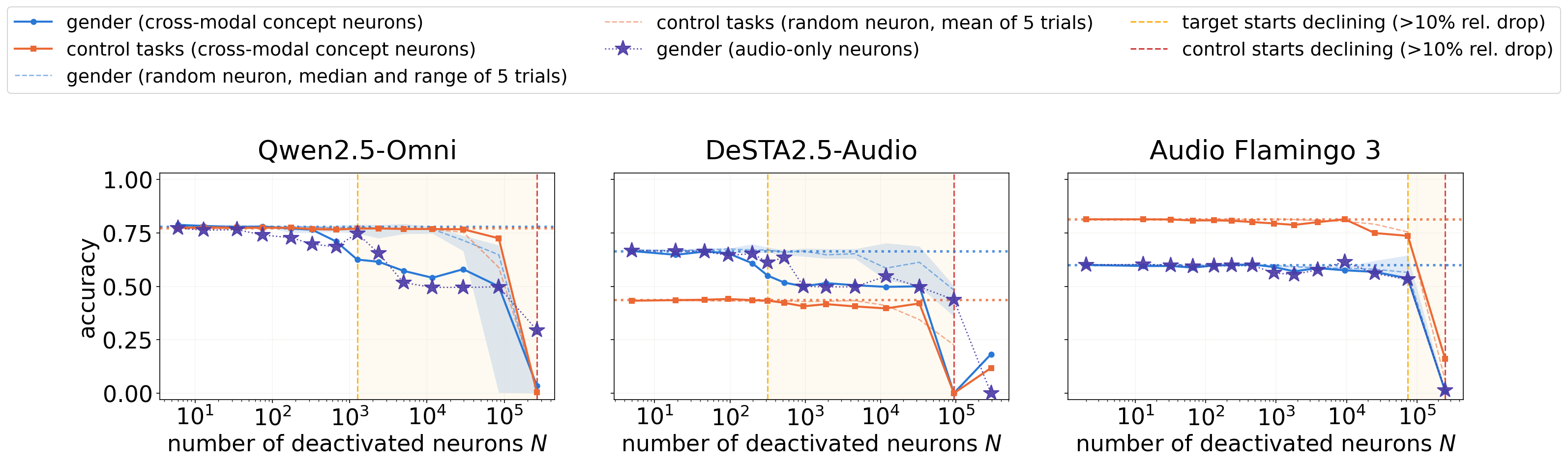}
  \caption{Task-level deactivation on the gender task, for Qwen2.5-Omni, DeSTA2.5-Audio, and Audio Flamingo 3 (left to right). Solid lines: target task accuracy (blue) and mean control-task accuracy (orange) as a function of the number of deactivated neurons $N$ (log scale). Dashed lines: matched-size random-neuron control (target task: median of the five random trials, with their min--max range shaded; control tasks: mean). Purple stars: audio-only neurons, a reference control (\S\ref{sec:exp3}). Vertical dashed lines and the shaded band mark the target's and controls' decline onsets; a band is shown only when the target's onset precedes the controls'.}
  \label{fig:exp3_gender}
\end{figure*}

\begin{figure*}[t]
  \centering
  \includegraphics[width=\textwidth]{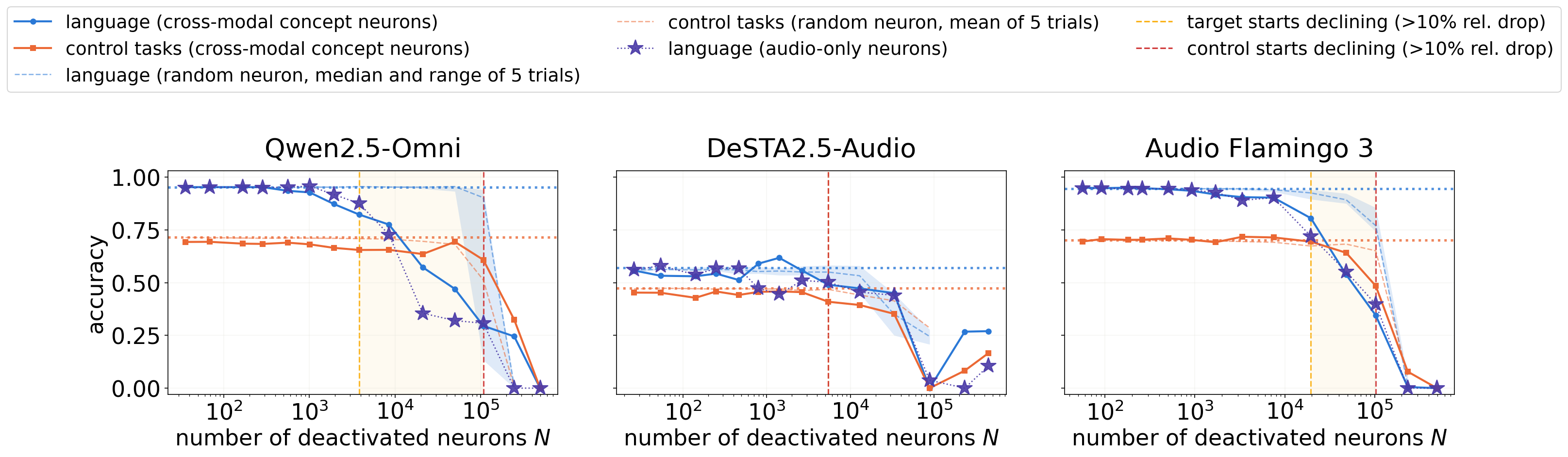}
  \caption{Task-level deactivation on the language task, for Qwen2.5-Omni, DeSTA2.5-Audio, and Audio Flamingo 3 (left to right). Solid lines: target task accuracy (blue) and mean control-task accuracy (orange) as a function of the number of deactivated neurons $N$ (log scale). Dashed lines: matched-size random-neuron control (target task: median of the five random trials, with their min--max range shaded; control tasks: mean). Purple stars: audio-only neurons, a reference control (\S\ref{sec:exp3}). Vertical dashed lines and the shaded band mark the target's and controls' decline onsets; a band is shown only when the target's onset precedes the controls'. Like animal, DeSTA2.5-Audio shows no selective window here.}
  \label{fig:exp3_language}
\end{figure*}

Figures~\ref{fig:exp3_animal}--\ref{fig:exp3_language} plot the same curves as Figure~\ref{fig:exp3_emotion} (\S\ref{sec:exp3}) for the remaining rows of Table~\ref{tab:exp3_necessity}: animal, gender, and language.

\section{Reproducibility and Artifact Details}

This appendix collects artifact, licensing, computational, and software details not included in the main text.

\subsection{Artifact licenses}
Qwen2.5-Omni-7B~\citep{xu2025qwen25omni} is released under the Apache-2.0 license. Audio Flamingo 3~\citep{goel2025audioflamingo3}'s code is released under the MIT license; its model weights are for non-commercial research use only. Neither DeSTA2.5-Audio~\citep{lu2025desta25audio} nor the SAKURA benchmark~\citep{yang2025sakura} states an explicit license on its public Hugging Face or GitHub page; we use both strictly for the same non-commercial academic research purpose as their respective release papers.

\subsection{Consistency with intended use}
All three models and the SAKURA benchmark are used exactly as released: as pretrained, frozen-at-inference LALMs and an off-the-shelf audio-attribute-recognition evaluation benchmark, respectively. We do not fine-tune, retrain, or redistribute any model weights or benchmark data; our steering and deactivation interventions (\S\ref{sec:exp4}, \S\ref{sec:exp3}) are applied only at inference time for scientific analysis, not as a proposed modification for deployment.

\subsection{Data screening}
We do not collect, curate, or redistribute any new data; all audio and labels used in this paper come unmodified from the public SAKURA benchmark (\S\ref{sec:setup}). We therefore did not perform additional screening for personally identifying or offensive content beyond whatever screening SAKURA's own creators already applied.

\subsection{Artifact documentation}
Each of the four SAKURA tasks we use provides 500 audio instances, covering every label it defines: emotion (100 instances per label; 5 labels: angry, disgust, fear, happy, sad), gender (250 per label; female, male), animal sound (9 species, 40--68 instances per label: cat, cow, dog, frog, and sheep at 68 each, crow, hen, pig, and rooster at 40 each), and spoken language (8 languages, 61--64 instances per label: English, French, German, and Spanish at 64 each, Chinese, Italian, Japanese, and Korean at 61 each). SAKURA does not publish speaker demographic information beyond what each task's own label encodes (e.g.\ binary gender); we did not have access to any further demographic or geographic characteristics of its speakers.

\subsection{Data statistics}
For every task, neuron selection and evaluation use a stratified 20\%/80\% validation/test split over these 500 instances (fixed seed 0; \S\ref{sec:method}), giving approximately 100 validation and 400 test instances per task, proportioned within each label. The overlap analyses (\S\ref{sec:exp2}--\S\ref{sec:exp1}) choose $k$ on the validation split and then reselect the neuron sets on the test split. Neuron sets used for interventions are always identified on the validation split, and every result reported in \S\ref{sec:results} is computed on the test split; the one choice that uses the test split is the steering neuron-set size (Appendix~\ref{sec:appendix_ksuff}; Limitations).

\subsection{Model size and computational budget}
Qwen2.5-Omni-7B's language backbone (its ``Thinker'' module) is based on Qwen2.5-7B (7B parameters); we do not use its speech-generating ``Talker'' module. DeSTA2.5-Audio's language backbone is Llama-3.1-8B-Instruct (8B parameters), frozen throughout its own training and throughout our experiments; its audio encoder is Whisper-large-v3. Audio Flamingo 3's language backbone is Qwen2.5-7B (7B parameters); its authors do not report a total parameter count including its AF-Whisper audio encoder and adapter. All experiments in this paper are inference-only: we never train or fine-tune any of the three models, only extract their activations or intervene on them (steering, deactivation) at inference time. Qwen2.5-Omni-7B and DeSTA2.5-Audio ran on RTX 3090 GPUs (24GB VRAM); Audio Flamingo 3 ran on RTX A6000 GPUs (48GB VRAM). Reproducing the experiments reported in this paper takes roughly 800 GPU-hours on single GPUs, most of it for the deactivation sweeps (\S\ref{sec:exp3}, Appendix~\ref{sec:appendix_ksuff}).

\subsection{Hyperparameter search and reproducibility}
\S\ref{sec:method} and \S\ref{sec:exp1}--\S\ref{sec:exp3} describe the neuron-set size $k$'s search grid (a doubling grid, extended for \S\ref{sec:exp3}) and selection criterion (Jaccard-maximizing on the validation split for \S\ref{sec:exp2}--\S\ref{sec:exp1}) at the point each is first used; Appendix~\ref{sec:appendix_ksuff} gives the one setting not otherwise specified there: the neuron-set size used for the steering experiment in \S\ref{sec:exp4}. All neuron selection and evaluation use the same fixed validation/test split (seed 0; described above) throughout.

\subsection{Descriptive statistics and single-run vs.\ averaged results}
All three models generate with greedy, temperature-free decoding (no sampling), so a given input and neuron intervention deterministically produces the same output; we do not report cross-seed generation variance for this reason. Three components of our pipeline are randomized and are reported accordingly: the Monte Carlo random-overlap baseline in Table~\ref{tab:appendix_exist_p} (100,000 samples); the five-trial random-neuron deactivation control in \S\ref{sec:exp3}, whose curves report the mean across trials; and the five random-neuron control sets per label in \S\ref{sec:exp4}, each drawn with a fixed seed; rather than averaging them, the significance tests use their spread as the null distribution (Appendix~\ref{sec:appendix_suff}). Every other number reported (accuracy, Jaccard overlap, cosine similarity) is a single deterministic evaluation pass over the fixed test split, not an average or a max over multiple runs.

\subsection{Existing software packages}
Our implementation uses PyTorch~\citep{paszke2019pytorch}, Hugging Face Transformers~\citep{wolf2020transformers}, SciPy~\citep{virtanen2020scipy} (for the hypergeometric and permutation, $t$, and rank significance tests in Appendix~\ref{sec:appendix_significance}), NumPy~\citep{harris2020array}, librosa~\citep{mcfee2015librosa} (for audio loading and resampling), and Matplotlib~\citep{hunter2007matplotlib} (for figures). Because DeSTA2.5-Audio and Audio Flamingo 3 require conflicting dependencies, each of the three models ran in its own environment: librosa (0.11.0) is common to all three; DeSTA2.5-Audio and Audio Flamingo 3 share PyTorch 2.10.0, SciPy 1.15.3, NumPy 2.2.6, and Matplotlib 3.10.8, while Qwen2.5-Omni instead used PyTorch 2.8.0, SciPy 1.16.3, NumPy 2.3.4, and Matplotlib 3.10.7.

\end{document}